\documentclass[prx,superscriptaddress,twocolumn,longbibliography]{revtex4-1}

\usepackage{graphicx}
\usepackage{amsmath}
\usepackage{amsfonts}
\usepackage{amssymb}
\usepackage{wasysym}

\usepackage{color,soul}

\begin{document}

\title{Chemical event chain model of coupled genetic oscillators}

\author{David J. J\"org}
\altaffiliation{Present address: Cavendish Laboratory, Department of Physics, University of Cambridge, JJ Thomson Avenue, Cambridge CB3 0HE, United Kingdom and The Wellcome Trust/Cancer Research UK Gurdon Institute, University of Cambridge, Tennis Court Road, Cambridge CB2 1QN, United Kingdom}
\affiliation{Max Planck Institute for the Physics of Complex Systems, N\"othnitzer Str.~38, 01187 Dresden, Germany}

\author{Luis G. Morelli}
\affiliation{Instituto de Investigaci\'on en Biomedicina de Buenos Aires (IBioBA)\---CONICET\---Partner Institute of the Max Planck Society, Polo Cient\'{\i}fico Tecnol\'ogico, Godoy Cruz 2390, C1425FQD, Buenos Aires, Argentina}
\affiliation{Departamento de F\'{\i}sica, FCEyN UBA, Ciudad Universitaria, 1428 Buenos Aires, Argentina}
\affiliation{Max Planck Institute for Molecular Physiology, Department of Systemic Cell Biology, Otto-Hahn-Str.~11, 44227 Dortmund, Germany}

\author{Frank J\"ulicher}\email{julicher@pks.mpg.de}
\affiliation{Max Planck Institute for the Physics of Complex Systems, N\"othnitzer Str.~38, 01187 Dresden, Germany}

\date{\today}

\begin{abstract}
\noindent%
We introduce a stochastic model of coupled genetic oscillators in which chains of chemical events involved in gene regulation and expression are represented as sequences of Poisson processes.
We characterize steady states by their frequency, their quality factor and their synchrony by the oscillator cross correlation.
The steady state is determined by coupling and exhibits stochastic transitions between different modes.
The interplay of stochasticity and nonlinearity leads to isolated regions in parameter space in which the coupled system works best as a biological pacemaker.
Key features of the stochastic oscillations can be captured by an effective model for phase oscillators that are coupled by signals with distributed delays.
\end{abstract}

\maketitle

\section{Introduction}

\begin{figure}[b]
\centerline{\includegraphics[width=8.6cm]{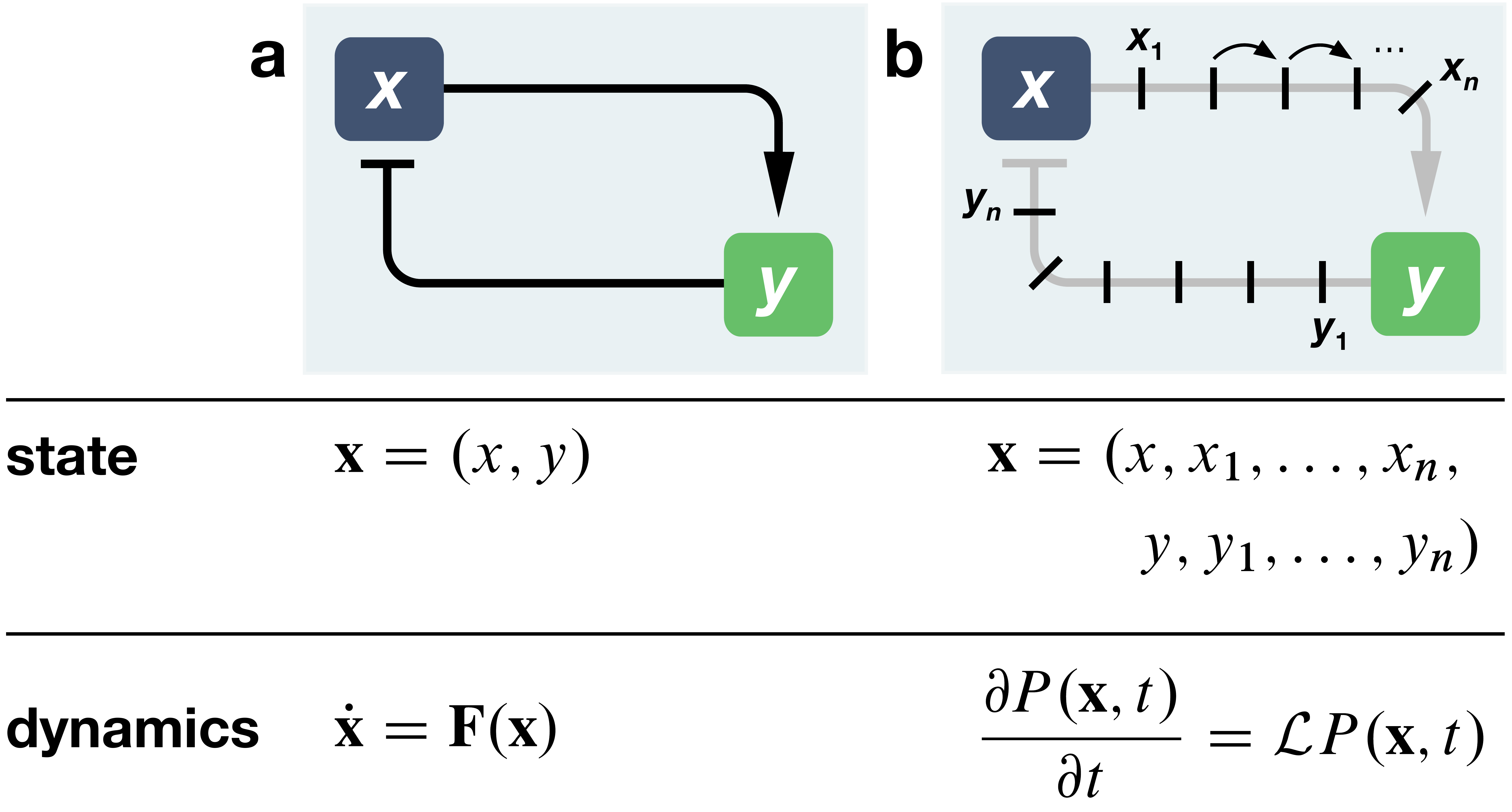}}
\caption{Exemplary depiction of a simple genetic feedback system with a gene $x$ and a gene product $y$. (a) Simplified feedback scheme that can be mapped onto a dynamic system and (b) representation as a chemical chain involving intermediate chemical states $x_i$ and $y_i$ that can be mapped onto a master equation.}	\label{fig.cec}
\end{figure}

\noindent%
Biological cells are complex dynamic systems which use specific proteins to activate and inhibit genes to ensure robust control of cellular functions \cite{alon,goldbeter}.
The production of such proteins themselves is mediated by gene activity, giving rise to feedback systems and complex dynamics.
The production of a gene product from an active gene comprises a series of chemical events such as transcription of DNA to RNA, splicing, and translation of RNA to a protein \cite{alberts}.
Gene regulation involves, e.g., the transport and binding of regulatory proteins to DNA.
Gene products can also serve as chemical signals that are mediated across different cells through so-called signaling pathways that involve production, transport, and binding of signaling molecules to receptor molecules or DNA \cite{ilagan07}.
Such sequences of chemical events typically involve the generation of intermediate products such as messenger RNA (mRNA), transcription factors, and signaling ligands.
Because of their complexity, such systems are often represented by simplified chemical rate equations that bypass intermediate steps and often neglect fluctuations \cite{novak08}. 
Motivated by an earlier approach \cite{morelli07}, we propose to describe genetic feedbacks by chemical event chains composed of a sequence of Poisson processes, see Fig.~\ref{fig.cec}.
These chains represent the sequences of transitions between intermediate chemical states, e.g., between mRNAs of different lengths and from mRNA to spliced mRNA.
This approach captures generic stochastic properties of complex cellular processes and can be used to generate an adequate stochastic description starting from a chemical reaction scheme.

Genetic oscillators are a prime example of genetic feedback systems, in which stochastic properties are important.
A prominent genetic oscillator is the circadian clock of humans, animals, and plants, where oscillations are used to provide information about the daytime to the organism \cite{hardin90,dunlap99,smolen02,schibler05,hastings07,zwicker10,zhang10b,abraham10,granada13}.
Genetic oscillators also play an important role during embryonic development, e.g., in neuronal differentiation \cite{imayoshi14,shimojo16b} and the segmentation of the body axis \cite{jorg15,oates12,shimojo16b}.
Genetic oscillators are characterized by gene regulatory networks that autonomously generate time-periodic changes in gene product numbers of so-called cyclic genes \cite{novak08,morelli07,niederholtmeyer15}.
This is typically achieved by a negative transcriptional feedback of the cyclic genes on themselves that involves a sufficiently large time delay \mbox{\cite{novak08,tokuda15,jorg17b}}.
In recent years, genetic oscillators have also been engineered in artificial systems \cite{elowitz00,garciaojalvo04,stricker08,danino10,kim11,niederholtmeyer15,potvintrottier16,tayar17}.
Both natural and artificial genetic oscillators exhibit pronounced amplitude and phase noise \cite{barkai00,vilar02,gonze02,mihalcescu04,garciaojalvo04,morelli07,zwicker10,potoyan14,webb16,lengyel17}, which limits their precision when used as a clock.
To achieve temporal and spatial coherence as well as high precision, cell-autonomous oscillators are typically coupled \cite{lewis03,liu07,oates12}. Such coupling facilitates synchronization and can affect the collective frequency \cite{needleman01,morelli09,herrgen10,ares12,cross12,shimojo16,liao16,jorg17,isomura17}.
Moreover, coupling between cellular oscillators via paracrine or juxtacrine signaling (i.e., via diffusible signals or contact-dependent signaling) typically proceeds at time scales similar to the oscillation period, implying the presence of coupling delays that can have profound effects on the coupled dynamics \mbox{\cite{morelli09,herrgen10,ares12,ananthasubramaniam14,tokuda15}}.

\begin{figure}[t]
\centerline{\includegraphics[width=8cm]{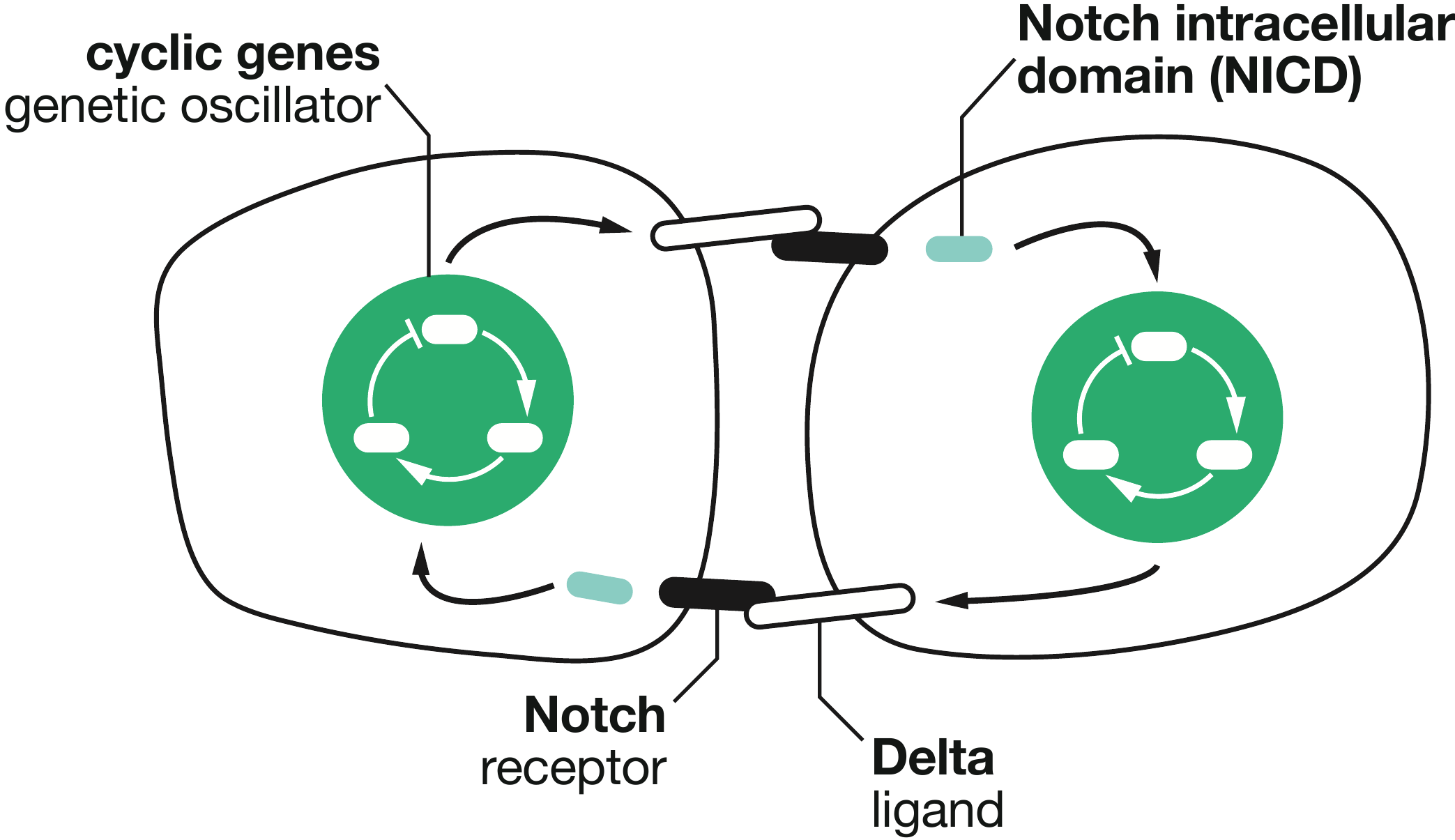}}
\caption{The zebrafish somitogenesis oscillator as an example for coupled genetic oscillations. Shown are two cells that act as autonomous oscillators and that are coupled through the Delta--Notch signaling pathway, an example for juxtacrine signaling~\cite{lewis03,lewis09}.
Coupling is bidirectional, that is, each cell acts as both a sender and a receiver.
During embryonic development, a tissue comprising these cellular oscillators guides the segmentation of the elongating body axis \cite{oates12}.
}\label{fig.zfo}
\end{figure}

In this paper, we present a framework to study the stochastic properties of genetic oscillators that are coupled by signaling pathways. As an example we consider the zebrafish somitogenesis oscillator  \cite{shimojo16b,lewis03}, see Fig.~\ref{fig.zfo}.
We investigate the precision and stochastic properties of collective oscillations and how they emerge from the kinetics of chains of chemical events.
Finally, we present an effective phase oscillator model that captures key features of this system. It is based on distributions of delay times in the oscillator coupling that captures collective frequency and stability properties of stochastic oscillations.

\section{Chemical event chain model of coupled genetic oscillators}
\begin{figure*}[t]
\centerline{\includegraphics[width=17.8cm]{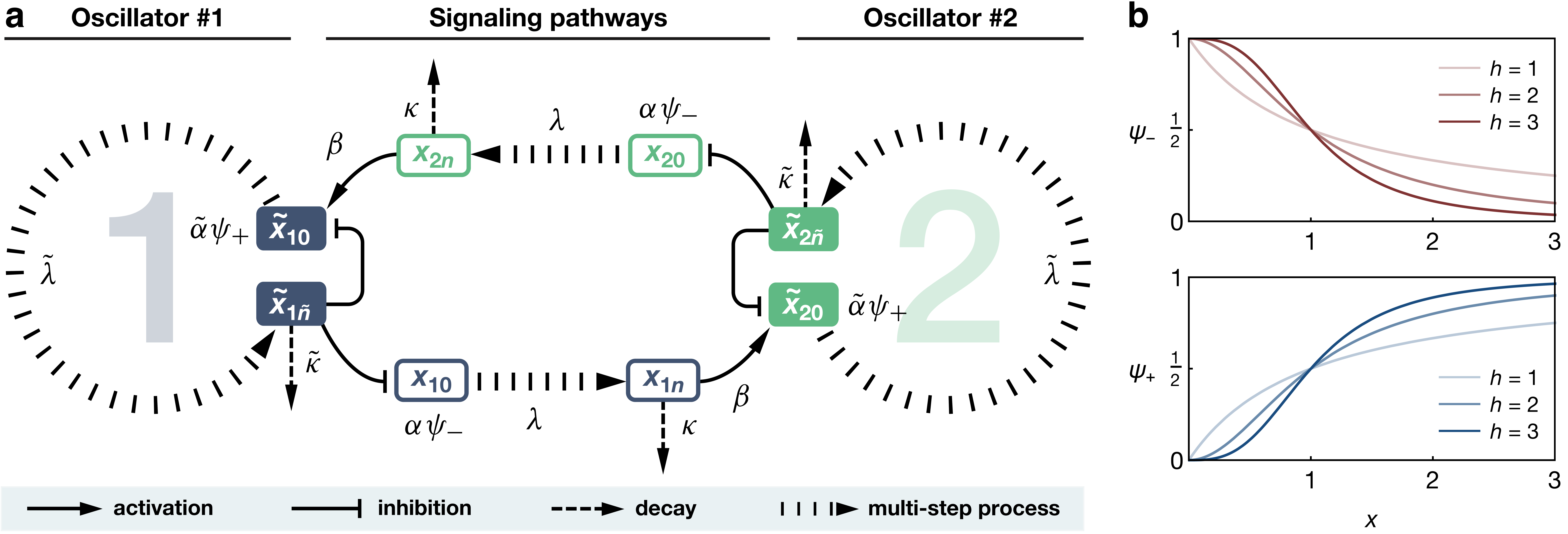}}
\caption{(a) Schematics of the chemical event chain model of two coupled genetic oscillators as described by Eqs.~(\ref{eq.me},\ref{eq.hill}). Boxes mark the initial and final products of a multi-step process and broken lines the intermediate products. (b) Hill functions $\psi_-$ and $\psi_+$ as given by Eqs.~(\ref{eq.hill}) for different values of the exponent $h$.}	\label{fig.scheme}
\end{figure*}

\noindent
We first introduce a Markov model for chemical event chains that captures the genetic interactions of coupled zebrafish somitogenesis oscillators (Fig.~\ref{fig.zfo}), see Fig.~\ref{fig.scheme}a.
The state of the system is specified by the occupation numbers of each step of the chain.
We denote the number of signaling molecules of oscillator $\mu =1,2$ at step $i=0,\hdots, n$ by ${x}_{\mu i}$ and the number of cyclic molecules at step $i = 0,\hdots,\tilde n$  by $\tilde{x}_{\mu i}$, see Fig.~\ref{fig.scheme}a. 
Synthesis of cyclic molecules takes place at the initial step $i = 0$ of the oscillators.
Molecules undergo a transition from step $i$ to $i+1$ at a constant transition rate and decay at the final step $i=\tilde n$.
The synthesis rate of both cyclic and signaling molecules is regulated by the amount of molecules at the final step $i=\tilde n$ of the oscillator.
Signaling molecules also undergo transitions through a sequence of steps with the last step of the signaling pathway regulating the synthesis rate of cyclic molecules in the receiving oscillator.

\subsection{Model formulation}

\noindent
We describe the dynamics of the system using a master equation~\cite{gardiner2009} that governs the time evolution of 
the probability $P(\mathbf{x},t)$ to find the system in the state $\mathbf{x}$ at time $t$, 
where
\begin{align*}
	\mathbf{x}=(\tilde{x}_{10},\hdots,\tilde{x}_{1\tilde{n}}, x_{10},\hdots,x_{1n}, \tilde{x}_{20},\hdots,\tilde{x}_{2\tilde{n}}, x_{20},\hdots,x_{2n})
\end{align*}
is the state vector of all occupation numbers.
The master equation is given by
\begin{widetext}%
\begin{align}%
\begin{split}%
\frac{\partial P}{\partial t} &= \sum_{\mu=1}^2 \Bigg\{ \underbrace{\tilde{\lambda} \sum_{i=0}^{\tilde{n}-1} \bigg[ (\tilde{x}_{\mu i}+1)\tilde{{\mathbb{E}}}^+_{\mu i} \tilde{{\mathbb{E}}}^-_{\mu, i+1} -  \tilde{x}_{\mu i} \bigg]}_{\text{oscillator chain}}
	 {} + {} \underbrace{\tilde\kappa \bigg[ (\tilde{x}_{\mu \tilde{n}}+1)\tilde{{\mathbb{E}}}_{\mu {n}}^+ - \tilde{x}_{\mu \tilde{n}} \bigg]}_{\text{decay of the cyclic product}} {} + {}  \underbrace{\psi_-\!\bigg(\frac{\tilde{x}_{\mu \tilde{n}}}{p}\bigg)\bigg[ \tilde\alpha + \beta \psi_+\!\bigg(\frac{{x}_{\bar\mu {n}}}{ q}\bigg) \bigg] (\tilde{{\mathbb{E}}}^-_{\mu 0}-1)}_{\text{regulation of cyclic genes}} \\
	 &\qquad\qquad {} + {} \underbrace{{\lambda} \sum_{i=0}^{{n}-1} \bigg[ ( x_{\mu i}+1) {{\mathbb{E}}}^+_{\mu i} {{\mathbb{E}}}
^-_{\mu, i+1} -  {x}_{\mu i} \bigg]}_{\text{signaling chain}} {} + {} 
	\underbrace{{\kappa} \bigg[ ( x_{\mu  n}+1){{\mathbb{E}}}^+_{\mu {n}} - {x}_{\mu {n}} \bigg]}_{\text{decay of the signaling product}}
 {} + {}  \underbrace{\alpha \psi_-\!\bigg( \frac{\tilde{x}_{\mu \tilde{n}}}{\tilde q} \bigg) ({{\mathbb{E}}}^-_{\mu 0}-1)}_{\text{regulation of signaling genes}} \Bigg\} P \ .
	\label{eq.me}
\end{split}
\end{align}
\end{widetext}
Here, $\bar\mu = 2 \delta_{\mu,1}+1 \delta_{\mu,2}$ refers to the index of the respective other oscillator. The creation and annihilation operators $\smash{\mathbb{E}^\pm_{\mu i}}$ increase or decrease the product levels $x_{\mu i}$ by one, 
$\smash{\mathbb{E}^\pm_{\mu i}} f(x_{10},\hdots, x_{\mu i},\hdots,x_{2 n})=f(x_{10},\hdots,x_{\mu i}\pm 1,\hdots,x_{2 n})$, 
and analogously for the operators $\smash{\tilde{\mathbb{E}}_{\mu i}}$ and product numbers $\tilde x_{\mu i}$ \cite{vanKampen07}.
The kinetic parameters characterizing the biochemical properties of gene expression and interaction are listed and explained in Table~\ref{table.parameters} and shown in Fig.~\ref{fig.scheme}a.
Activation and repression of gene expression at the initial stages of the oscillators and the signaling pathways are described by functions of the Hill type~\cite{novak08},
\begin{align}
	\psi_-(x) = \frac{1}{1+x^h} \ ,  \qquad
	\psi_+(x) = \frac{x^h}{1+x^h} \ , \label{eq.hill}
\end{align} 
where $\psi_-$ describes inhibition and $\psi_+$ describes activation and the exponent $h$ determines the nonlinearity of the feedback, see Fig.~\ref{fig.scheme}b.
Here, we are interested in steady-state solutions of the master equation~(\ref{eq.me}), which describe the long-term collective behavior of the system.

\subsection{Characterization of oscillator coupling via stochastic signaling}
\label{subsec:coupling}

\noindent
We first summarize features of the introduced coupling process that are useful to parametrize the system and understand its limiting cases.
The coupling strength depends on several model parameters: the production rate $\alpha$ and inhibition threshold $\tilde q$ of the signaling molecules in the sending oscillator, as well as the activation rate $\beta$ and the activation threshold~$q$ in the receiving oscillator, see Fig.~\ref{fig.scheme} and Table~\ref{table.parameters}.
The limiting case of uncoupled oscillators can be realized through either (i) $\alpha=0$, (ii) $\beta=0$, (iii) $q\to\infty$, (iv) $\tilde q \to 0$, or any combination thereof.

Moreover, the chain of Poisson processes of the signaling pathway effectively generates 
a Gamma distribution of arrival times for molecules starting at step $i=0$ and arriving at step $i=n$ \cite{morelli07}, 
\begin{align}
	g(t) = \frac{\lambda^{ n}}{( n-1)!}  t^{ n-1} \mathrm{e}^{- \lambda t} \ .\label{eq.gamma}
\end{align}%
Hence, the mean $\tau$ and variance $\sigma^2$ of this distribution characterize the mean signaling delay and the dispersion of signaling delays,
\begin{align}
	\tau =  n/\lambda \ , \qquad \sigma^2 = n/\lambda^2 \ . \label{eq.tau}
\end{align}
The arrival time distribution $g$ can be interpreted as a memory kernel for a probability that effectively summarizes the effects of noise and delays introduced by stochastic signaling.

\subsection{Correlation functions and oscillator quality} \noindent
\label{subsec:correlation}

\noindent %
Before studying the dynamics of the coupled genetic oscillator system, we introduce measures that characterize their function: the quality factor measuring frequency fluctuations and the cross correlation measuring synchrony.
To define these quantities, we introduce the temporal correlation function~$C_{\mu\nu}$ between the 
final products~$\tilde{x}_{\mu \tilde{n}}$ and $\tilde{x}_{\nu \tilde{n}}$ of the oscillators~$\mu$ and $\nu$,
\begin{align}
	C_{\mu\nu}(t) = \langle \tilde{x}_{\mu \tilde{n}}(t) \tilde{x}_{\nu \tilde{n}}(0) \rangle - \langle \tilde{x}_{\mu \tilde{n}} \rangle \langle \tilde{x}_{\nu \tilde{n}} \rangle \ , \label{eq.crosscorrelation.function}
\end{align}
where the brackets denote steady-state expectation values.
The quality factor of an oscillator can be defined as follows.
The normalized temporal autocorrelation function of an oscillator is given by
\begin{align}
	G(t) =  \frac{C_{\mu\mu}(t)}{C_{\mu\mu}(0)}  \ . \label{eq.autocorrelation}
\end{align}
Since both oscillators and signaling pathways are entirely identical, the autocorrelation $G$ does not depend on the oscillator index $\mu$.
For a noisy oscillator, this autocorrelation function typically exhibits a functional form of the type
$G(t) \simeq \cos(2\pi t/T) \mathrm{e}^{-t/t_\mathrm{c}}$ for large $t$,
where $T$ is the period of oscillations and $t_\mathrm{c}$ is the correlation time.
We define the quality factor~$Q$ as the dimensionless ratio of the correlation time and the oscillation period~\cite{morelli07,stratonovich63},
\begin{align}
	Q = t_\mathrm{c}/T \ . \label{eq.qualityfactor}
\end{align}
The quality factor $Q$ corresponds to the number of cycles over which the oscillatory signal stays highly correlated, 
thus quantifying the number of cycles over which the oscillators serve as a viable clock.

\begin{figure}[t]
\centerline{\includegraphics[width=7.5cm]{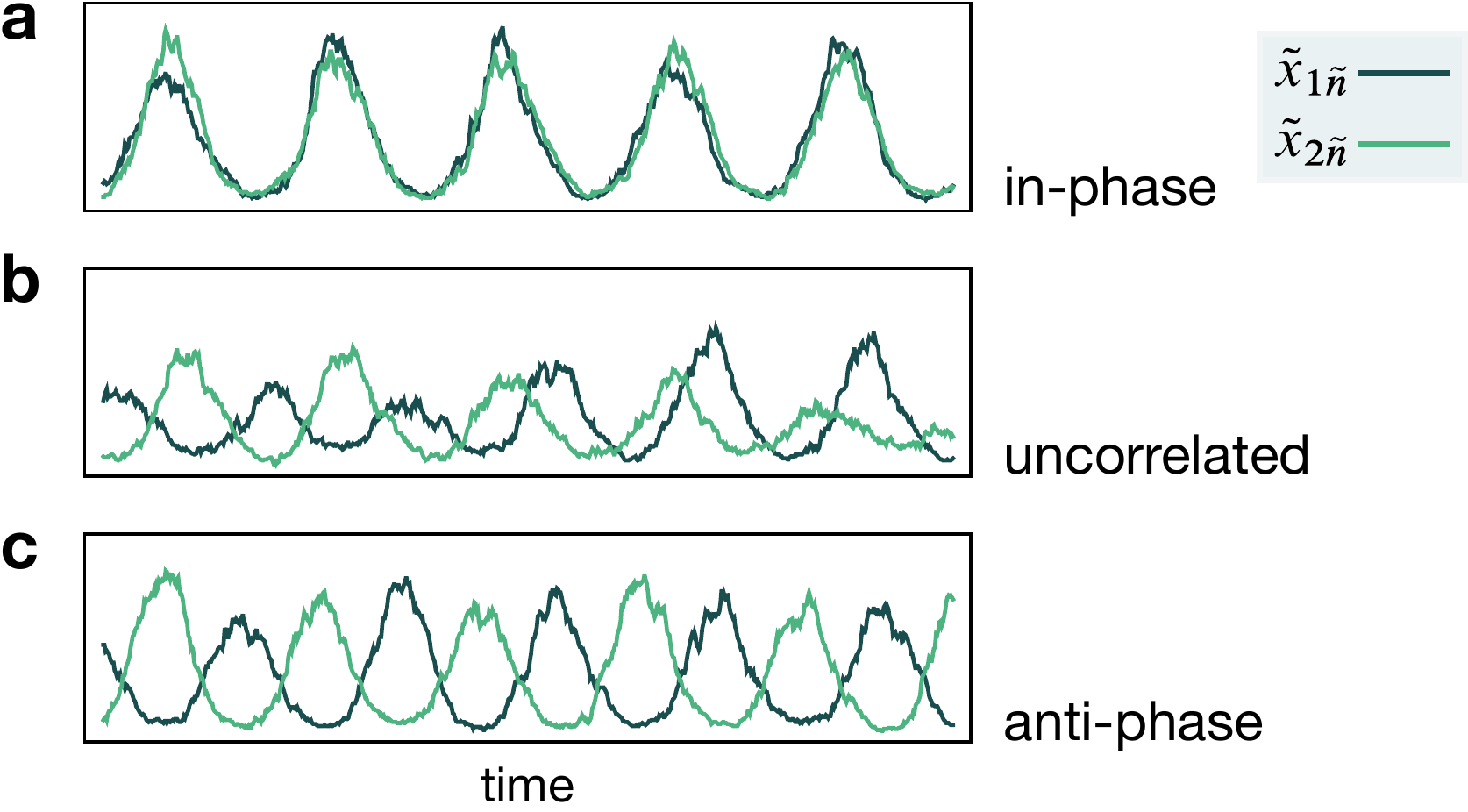}}
\caption{Example time series of the cyclic final products~$\tilde x_{1 \tilde n}$ and $\tilde x_{2 \tilde n}$ for different synchronization scenarios: (a) in-phase, (b) uncorrelated, and (c) anti-phase. For all plots, parameters are given in Table~\ref{table.parameters} except for the transition rate $\lambda$, which is chosen such that the effective coupling delay $\tau$, given by Eq.~(\ref{eq.tau}), takes values $\tau=0.1T$ (a), $\tau=0.2T$ (b), and $\tau=0.5T$ (c), where $T=28$ is the period of the uncoupled oscillators.
}
\label{fig.precisionandsync}
\end{figure}

The synchrony of two stochastic trajectories is related to the degree of correlation of their individual dynamics. 
To quantify synchrony, we compute the normalized cross-correlation of the final products of the oscillators,
\begin{align}
	C = \frac{C_{12}(0)}{\sqrt{C_{11}(0)C_{22}(0)}} \ ,\label{eq.crosscorrelation}
\end{align}
with $C_{\mu\nu}$ given by Eq.~(\ref{eq.crosscorrelation.function}).
The cross-correlation~$C$ describes the fraction of shared fluctuations between both signals and its sign indicates the mode of synchrony: $C$ takes values in the interval $[-1,1]$, ranging from perfect correlation ($C=1$) to no correlation ($C=0$) to perfect anti-correlation ($C=-1$), which in the case of oscillations corresponds to in-phase oscillations, phase-drifting oscillations and anti-phase oscillations, see Fig.~\ref{fig.precisionandsync}.

\section{Frequency, quality and synchrony of coupled genetic oscillators}
\noindent We numerically investigate frequency, quality, and synchrony of the coupled genetic oscillators by generating multiple realizations of the stochastic process described by the master equation (\ref{eq.me}) using numerical simulations, see Fig.~\ref{fig.precisionandsync} for examples. The simulation method is detailed in Appendix~\ref{sec:simulations}.

\squeezetable
\begin{table}[t]
\begin{ruledtabular}
\begin{tabular}{cccl}
 Param. & {Unit} & Value & Description \\ \hline
\multicolumn{4}{l}{Oscillators}\\
	$\tilde{n}$ & 1 & 18 & number of steps\\
	$\tilde\alpha$ & $\mathrm{N}\mathrm{T}^{-1}$ & 60 & basal production rate \\
	$\tilde\lambda$ & $\mathrm{T}^{-1}$ & 1.5 & transition rate between steps \\
	$\tilde\kappa$ & $\mathrm{T}^{-1}$ & 0.5 & decay rate for the final product \\ 
		$p$ & $\mathrm{N}$ & 20 & threshold for cyclic auto-inhibition \\
	$\beta$ & $\mathrm{N}\mathrm{T}^{-1}$ & 20 & activation strength due to signaling \\
	\hline
\multicolumn{4}{l}{Signaling pathways}\\
	$n$ &  1 & 10 &  number of steps \\
	$\alpha$ & $\mathrm{N}\mathrm{T}^{-1}$ & 60 & basal production rate \\
	$\lambda$ & $\mathrm{T}^{-1}$ & 0.5 & transition rate between steps \\
	$\kappa$ & $\mathrm{T}^{-1}$ & 0.5 & decay rate for the final product \\ 
	$q$ & $\mathrm{N}$ & 100 & threshold for activation by signaling \\ 
	$\tilde q$ & $\mathrm{N}$ & 20 & threshold for repression of signaling \\ \hline
	$h$ & 1 & 2 & Hill exponent (repression, activation) \\
\end{tabular}
\end{ruledtabular}
\caption{Parameters used for numerical simulations. `N' refers to molecule numbers and `T' is the unit of time.}
\label{table.parameters}
\end{table}%

\begin{figure*}[t]
\centerline{\includegraphics[width=15cm]{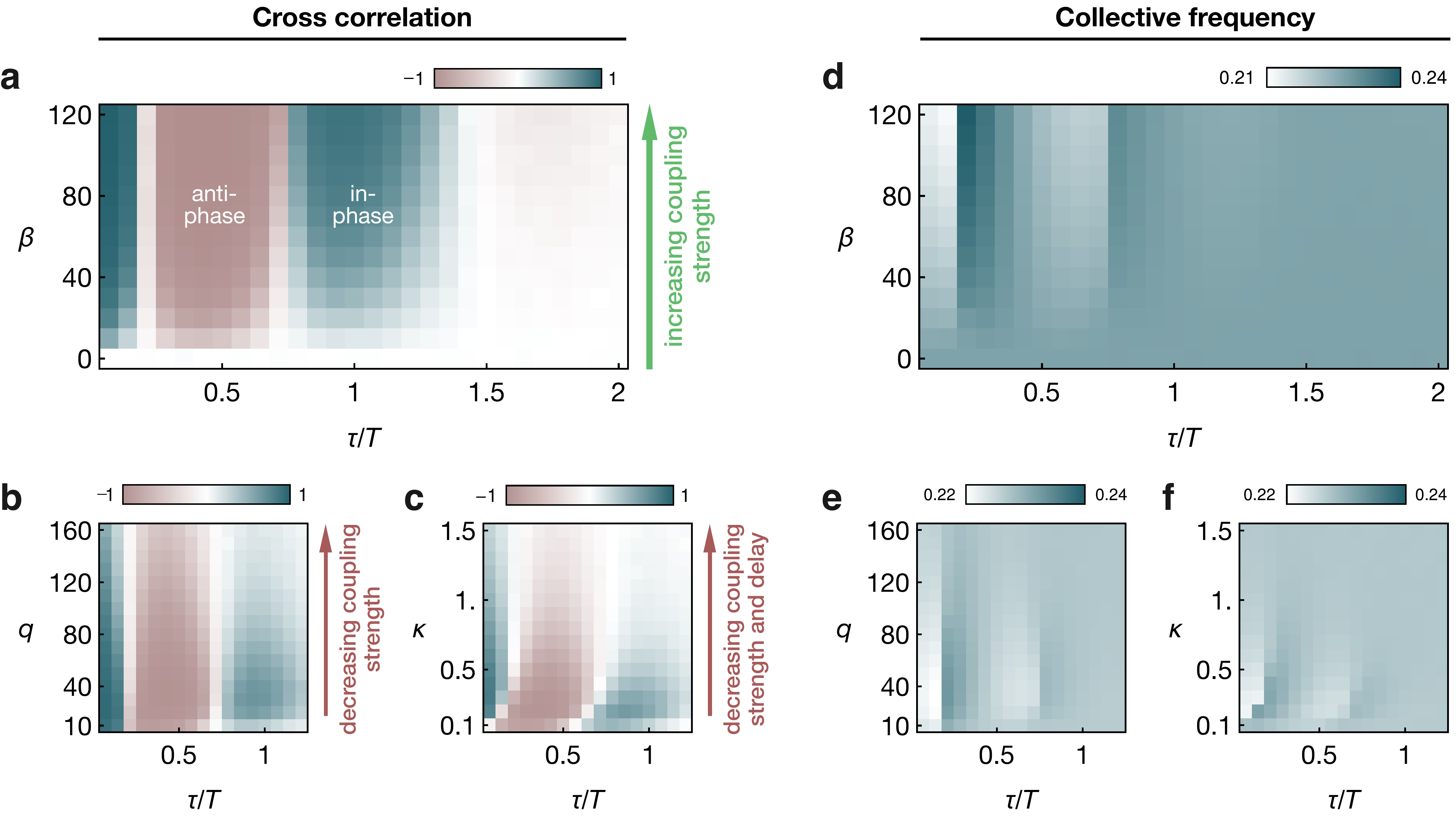}}
\caption{Coupling delays determine the mode of synchrony and the collective frequency. (a--c) Density plots of the cross correlation $C$, Eq.~(\ref{eq.crosscorrelation}). Blue colors indicate positive values of $C$ corresponding to in-phase correlations, red colors indicate negative values of $C$ corresponding to anti-phase correlations. (d--f) Density plots of the collective frequency $\Omega$. The parameters that are not varied are given in Table~\ref{table.parameters}.}
\label{fig.modes}
\end{figure*}

\subsection{Coupling delays determine the mode of synchrony}	
\label{subsec:delays}
\noindent
First, we study how coupling via stochastic event chains affects the mode of synchrony of the two oscillators.
To this end, we focus on the parameters $\beta$ and $q$ as a measure of coupling strength and vary the mean signaling delay $\tau$ by changing the transition rate $\lambda$ (see Section~\ref{subsec:coupling}).
We find that stochastic signaling delays determine whether the oscillator system exhibits in-phase, anti-phase, or uncorrelated oscillations:
Figs.~\ref{fig.modes}a--c show density plots of the cross correlation $C$ as a function of the effective coupling delay $\tau$ as well as the activation rate $\beta$, the activation threshold $q$, and the decay rate $\kappa$ of the signaling molecules.
These plots reveal an alternation of in-phase and anti-phase correlated regions as a function of the signaling delay~$\tau$ with in-phase regions located around integer multiples of the uncoupled period $T$ and anti-phase regions located around odd multiples of $T/2$.
This behavior is generally known for coupled oscillators with delayed coupling~\cite{schuster89}.
For increasing signaling delays~$\tau$, which imply increasing dispersions $\sigma$ of delay times in our parametrization, the correlations between the oscillators decay until eventually, for large signaling delays, the oscillators become effectively uncoupled.
Regions with a high degree of correlation (large $|C|$) are separated by regions of uncorrelated oscillations.

\subsection{Coupling affects the collective frequency}	
\begin{figure}[t]
\centerline{\includegraphics[width=6.5cm]{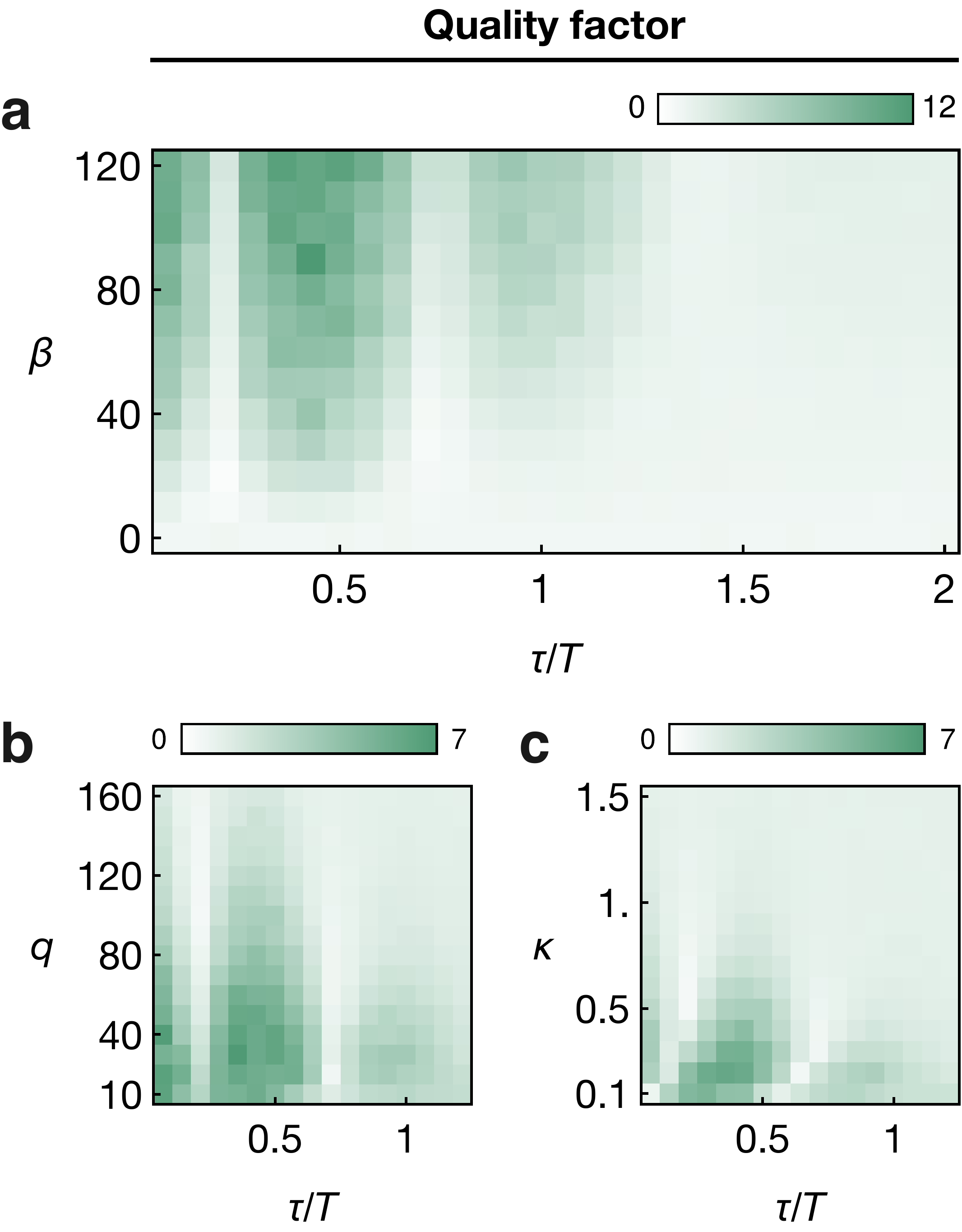}}
\caption{Coupling enhances precision. Density plots of the quality factor $Q$, Eq.~(\ref{eq.qualityfactor}), 
as a function of the scaled mean coupling delay $\tau / T$ and 
(a) the activation strength $\beta$, 
(b) the activation threshold $q$, and 
(c) the decay rate~$\kappa$ of signaling molecules.
The graded bar on top of each panel indicates the range of $Q/Q_0$ values where $Q_0 = 20.8$ is the quality of an uncoupled oscillator.
The parameters that are not varied are given in Table~\ref{table.parameters}.
}
\label{fig.quality}
\end{figure}

\noindent
As shown above, coupling tends to synchronize oscillators, implying that they attain a common collective frequency.
If coupling is delayed, as is the case here, this collective frequency can differ from the frequency of the uncoupled oscillators~\cite{schuster89,yeung99}.
Figs.~\ref{fig.modes}d--f show density plots of the collective frequency~$\Omega$ of both oscillators, obtained from the autocorrelation, Eq.~(\ref{eq.autocorrelation}), as a function of the same parameters as in Figs.~\ref{fig.quality} and \ref{fig.modes}.
As a function of the signaling delay~$\tau$, these plots reveal sharp changes of the frequency at odd multiples of $T/4$.
This indicates that the collective frequency of in-phase correlated states is distinct from those of anti-phase correlated states, compare to Figs.~\ref{fig.modes}a--c.
For large signaling delays, the effect of coupling on the collective frequency vanishes.
We will address the dependence of the collective frequency on the signaling delay when studying a phase oscillator approximation in Section~\ref{sec:phase.oscillators}.

\subsection{Coupling enhances the quality of oscillations}

\noindent Next we investigate how the precision of oscillations in the coupled system is affected by stochastic coupling.
Fig.~\ref{fig.quality} shows the quality factor $Q$ as a function of the same parameters as in Fig.~\ref{fig.modes}.
As the delay $\tau$ increases, the quality factor shows distinct maxima and minima of decreasing magnitude.
Maxima of the quality appear in region where both oscillators show a high degree of in-phase or anti-phase correlation, compare to Figs.~\ref{fig.modes}a--c.
For large delays, the quality settles towards a low value and eventually becomes independent of the delay.
This decay of the quality is due to the increase in the width $\sigma$ of the delay distribution which accompanies the increase in $\tau$, see Eq.~(\ref{eq.tau}). 
As the spread of the delay distribution increases, temporal information about the sending oscillator's state is lost along the signal pathway due to fluctuations and thus, information transmission becomes unreliable.

The coupling strength affects the quality factor as well, with stronger coupling leading to higher qualities:
The quality factor monotonically increases with the activation strength $\beta$.
In the parameter range investigated here, coupling can lead to  an order of magnitude increase of the quality compared to the uncoupled case, see Fig.~\ref{fig.quality}a.
The activation threshold~$q$ has a qualitatively different effect on $Q$.
Quality is maximized for threshold levels~$q$ close to the mean final product number~$\langle {x}_{\mu {n}}\rangle$ of the signaling pathways and decays for smaller or larger values, see Fig.~\ref{fig.quality}b.
Interestingly, since the quality depends on the mean signaling delay~$\tau$ in the same non-monotonic way as above, islands of high quality containing local extrema are observed in parameter space.
The behavior of the quality as a function of the decay rate~$\kappa$ as shown in Fig.~\ref{fig.quality}c suggests that in addition to the signaling delay~$\tau$, the decay time $\kappa^{-1}$ of the signaling molecules effectively contributes to the total coupling delay.
This is indicated by the leftward tilt of high quality islands for low decay rates, see Fig.~\ref{fig.quality}c.
Again, for constant decay rate~$\kappa$, a non-monotonic behavior of the quality in the coupling delay~$\tau$ can be observed.
Hence, we find that the non-monotonic behavior of the quality in both the coupling delay and the coupling strength results in multiple separated islands in parameter space that give rise to  high precision oscillations.

\begin{figure*}[t]
\centerline{\includegraphics[width=13cm]{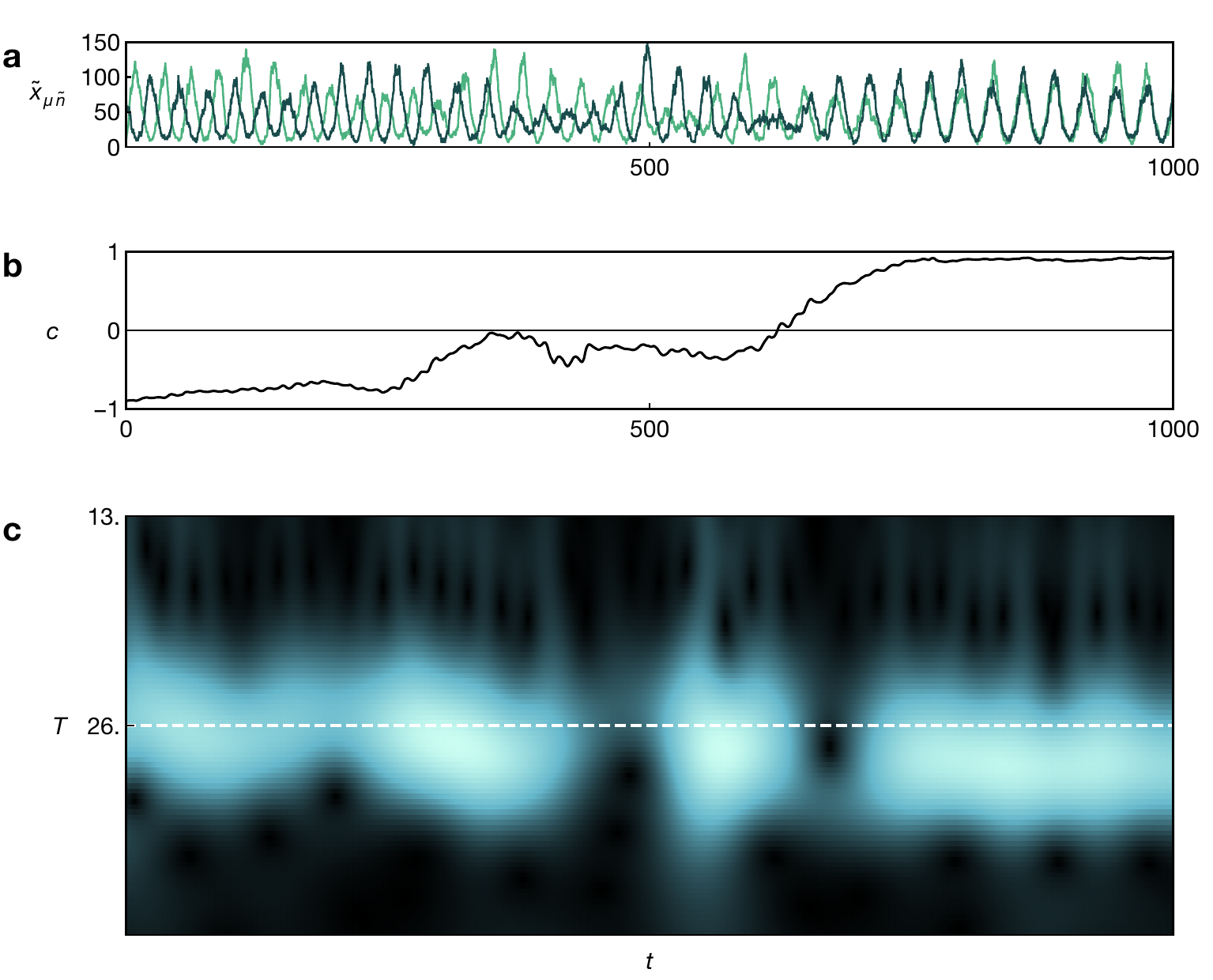}}
\caption{Stochastic switching between anti-phase and in-phase synchronized oscillations. 
(a) Time series of products $\tilde x_{1\tilde n}$ and $\tilde x_{2\tilde n}$ showing a switching event between anti-phase and in-phase oscillations.
Parameters are given in Table~\ref{table.parameters} except for $\beta=120$ and $\lambda=5/3$.
(b) Windowed cross correlation~$c$ of the time series in A, see Eq.~(\ref{eq.windows.crosscorrelation}), showing the three distinct regimes. The width of the averaging window is $w=4T$.
(c) Wavelet scalogram of one of the time series in A. 
Bright regions indicate strong period components.
The white dashed line serves as a guide for the eye.}
\label{fig.switching}
\end{figure*}

\subsection{Stochastic switching between in-phase and anti-phase synchrony}
\label{subsec:switching}
\noindent
In the transition regions between in-phase and anti-phase correlations (where $\tau \approx (2n+1) T/4$ with integer $n$ and $T$ being the period of the uncoupled oscillators), stochastic switching between in-phase and anti-phase correlated oscillations occurs in single realizations of the system.
Switching events can be observed by direct inspection of the time series of final products, see Fig.~\ref{fig.switching}a.
The degree of correlation within a single realization of the system can be displayed by means of the windowed normalized cross correlation,
\begin{align}
	c(t) = \frac{c_{12}(t)}{\sqrt{c_{11}(t)c_{22}(t)}} \ , \label{eq.windows.crosscorrelation}
\end{align}
where
\begin{align}
	c_{\mu\nu}(t) = \langle\!\langle \tilde{x}_{\mu \tilde{n}} \tilde{x}_{\nu \tilde{n}} \rangle\!\rangle_t - \langle\!\langle \tilde{x}_{\mu \tilde{n}} \rangle\!\rangle_t \langle\!\langle \tilde{x}_{\nu \tilde{n}} \rangle\!\rangle_t\ ,
\end{align}
and $\langle\!\langle f \rangle\!\rangle_t = w^{-1} \int_{-w/2}^{w/2} f(t+s)\,\mathrm{d}s$ denotes a time average over a time window with width $w$.
Fig.~\ref{fig.switching}b shows the windowed cross correlation $c$ for the realization shown in A.
Starting from anti-phase correlations ($c \approx -1$), the system goes through an extended phase of uncorrelated oscillations before attaining an in-phase synchronized state ($c \approx 1$).
With the mode of synchrony, the collective frequency of the system changes as well:
Fig.~\ref{fig.switching}c shows a density plot of a wavelet transform of one of the time series, corresponding to a time-dependent period spectrum (see Appendix~\ref{sec:wavelet.transform} for technical details).
During the anti-phase state, the bright regions, indicating strong period components, are centered around the white dashed line.
After the transition to the in-phase state, the bright regions fall almost entirely below the white dashed line, indicating an increased period.
This frequency change is consistent with the frequency differences between in-phase and anti-phase correlated states, compare to Figs.~\ref{fig.modes}d--f.

Clearly, stochastic switching between different modes of synchrony and collective frequencies including extended transient periods affects the long-time behavior of the autocorrelation, effectively resulting in an impairment of the precision.
In addition, the presence of two slightly detuned collective frequencies leads to beating patterns in the autocorrelation function Eq.~({\ref{eq.autocorrelation}}), so that in these cases, the quality factor $Q$ obtained from fits of the autocorrelation captures the average period and an effective correlation time (see Section~{\ref{subsec:correlation}}) and can even drop below the single-oscillator quality $Q_0$.
This impairment contributes to the low quality regimes that separate the regions of high quality observed in Fig.~\ref{fig.quality}.

\section{Effective phase model}
\label{sec:phase.oscillators}
\noindent
The chemical event chain model Eq.~(\ref{eq.me})
describes how stochastic coupling affects the collective modes and their frequency, see Fig.~\ref{fig.modes}.
We aim to capture the key features of these collective modes using a simpler theory of delay-coupled phase oscillators.
Phase oscillator models reduce the complexity of limit cycle oscillators to the dynamics of a phase variable $\phi \in [0,2\pi)$ representing the state of oscillator while neglecting the amplitude dynamics \cite{kuramoto84,acebron05,rodrigues16}.

\subsection{Phase oscillators with distributed coupling delay times}

\noindent
The dynamics for the phase $\phi_\mu$ of oscillator~$\mu=1,2$ is given by
\begin{align}
	\frac{\mathrm{d}\phi_\mu}{\mathrm{d}t} = \omega +  \varepsilon \int_0^\infty g(s) \sin(\phi_{\bar\mu}(t-s) - \phi_\mu(t) ) \, \mathrm{d}s \ ,\label{stos.phaseoscillators}
\end{align}
where $\omega$ is the intrinsic frequency of the autonomous oscillators, $\varepsilon$ is the coupling strength, $g$ is the distribution of delay times, and $\bar \mu$ denotes the respective other oscillator as in Eq.~(\ref{eq.me}).
The coupling term in Eq.~(\ref{stos.phaseoscillators}) dynamically alters the instantaneous frequency of the oscillator depending on the phase relationship to the other oscillator.
For the distribution $g$ of delay times, we here choose the Gamma distribution Eq.~(\ref{eq.gamma}) that describes the distribution of arrival times of signaling molecules.

We now show that Eq.~(\ref{stos.phaseoscillators}) can describe many qualitative features of the in-phase and anti-phase synchronized states observed in the stochastic theory.
The in-phase synchronized state, given by $\phi_\mu(t) = \Omega t$,
is characterized by both oscillators evolving with the same collective frequency~$\Omega$ and having no phase lag relative to each other.
Using this ansatz in Eq.~(\ref{stos.phaseoscillators}) yields an implicit transcendental equation for $\Omega$,
\begin{align}
\begin{split}
	\Omega &= \omega - \varepsilon \left( \frac{1}{1+\Omega^2/   \lambda^2} \right)^{n/2} \sin \bigg(   n \arctan \frac{\Omega}{  \lambda} \bigg) \ .\label{eq.collectivefrequency}
\end{split}
\end{align}
The system can also exhibit an anti-phase synchronized state, $\phi_1(t) = \bar\Omega t = \phi_2(t) + \pi$, where the corresponding collective frequency $\bar\Omega$ obeys Eq.~(\ref{eq.collectivefrequency}) with $\varepsilon$ replaced by $-\varepsilon$.
In both cases, the collective frequency satisfies the bound
\begin{align}
	\omega - \varepsilon \leq \Omega \leq \omega+\varepsilon \ ,
\end{align}
implying that $\Omega$ can only deviate from the intrinsic frequency~$\omega$ by the coupling strength $\varepsilon$. Moreover, we find that $\Omega\to\omega$ for $\lambda\to 0$, implying that for increasing delays due to an increased jump rate, the collective frequency becomes independent of coupling, a behavior that we also found in the chemical event chain model, see Fig.~\ref{fig.modes}d--f.
The special case of a single discrete delay, $g(t)=\delta(t-\tau)$, corresponds to the limit $n\to\infty$ with $\lambda=n/\tau$ and fixed $\tau$ and in this case, Eq.~(\ref{eq.collectivefrequency}) becomes 
\begin{align}
	\Omega = \omega - \varepsilon \sin (\Omega\tau) \ , \label{eq.collectivefrequency.discrete}
\end{align}
a result well-known in the literature \cite{schuster89,yeung99,earl03,ares12}. 

\begin{figure}[t]
\centerline{\includegraphics[width=7.5cm]{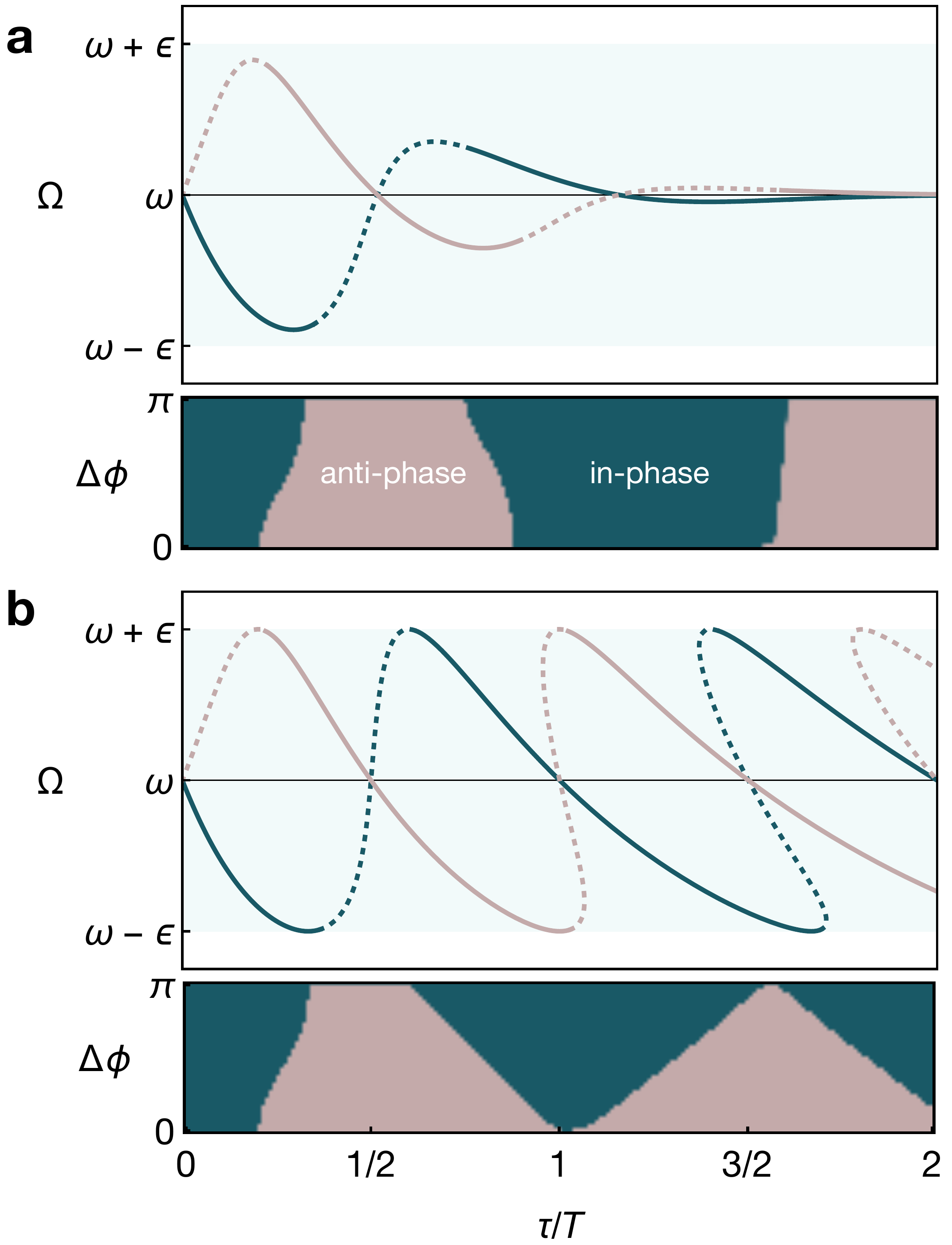}}
\caption{
Collective frequency~$\Omega$ of the phase oscillator system Eq.~({\ref{stos.phaseoscillators}}) (curves) and its asymptotic state for given constant initial phase differences $\Delta\phi$ (region plots) as a function of the scaled coupling delay~$\tau/T$ (see Section~{\ref{subsec:comparison}}). Blue curves show the in-phase synchronized state, red curves show the anti-phase synchronized state. Solid curves show stable solutions, dashed curves show unstable solutions.
(a) System with a distribution of delay times for $n=10$ and different mean delays $\tau$ obtained by varying $\lambda$. The collective frequency is determined by Eq.~({\ref{eq.collectivefrequency}}). (b) Collective frequency for a discrete delay time $\tau$, determined by Eq.~({\ref{eq.collectivefrequency.discrete}}).
In both plots, $\omega=0.224$ and $\varepsilon=\omega/4$.}
\label{fig.phase.oscillator.frequency}
\end{figure}

Fig.~\ref{fig.phase.oscillator.frequency} shows the collective frequencies for systems with a distribution of delays, Eq.~(\ref{eq.collectivefrequency}), and for systems with a discrete coupling delay, Eq.~(\ref{eq.collectivefrequency.discrete}).
In both cases, for a given set of parameters, this equation can exhibit multiple solutions in $\Omega$.
However, compared to a discrete delay, a distribution of delay times leads to a decaying dependence of the collective frequency on the coupling delay if the number of steps $n$ is fixed.
Using a standard linear stability analysis, an analytical criterion for the stability of the in-phase and anti-phase synchronized states can be found, see Appendix~\ref{sec:linear.stability}.
In Fig.~\ref{fig.phase.oscillator.frequency}, stable states are indicated by solid curves, unstable states by dashed curves.
This also illustrates that in-phase and anti-phase solutions can be simultaneously stable in certain parameter regions.

\begin{figure*}[t]
\centerline{\includegraphics[width=18.1cm]{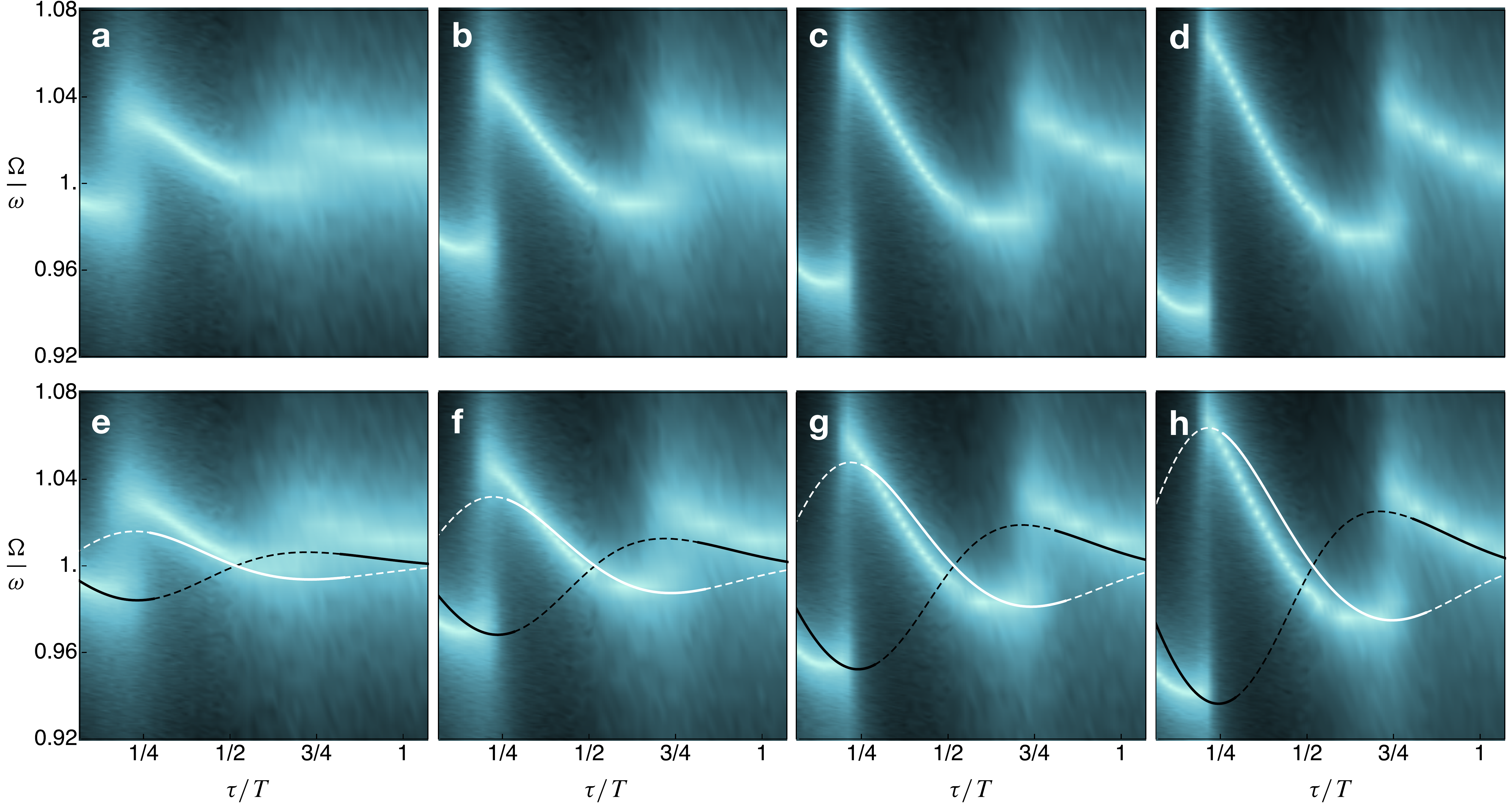}}
\caption{Collective frequency as a function of the scaled coupling delay $\tau/T$ for different activation strengths $\beta$: (a, e)~$\beta=30$, (b, f)~$\beta=60$, (c, g)~$\beta=90$, (d, h)~$\beta=120$.
(a--d) The density plots show logarithmic periodograms of oscillations in the chemical event chain model, where bright regions correspond to strong frequency components.
(e--h) The curves (superimposed on the same density plots as in a--d) show the collective frequencies of the in-phase (black) and anti-phase (white) solutions of the phase model, Eq.~(\ref{eq.collectivefrequency}). Solid lines show stable solutions, dashed lines show unstable solutions. The delay is given in multiples of the uncoupled period $T$, the collective frequency $\Omega$ is given in multiples of the uncoupled frequency $\omega=2\pi/T$. Solid lines indicate stable solutions, dashed lines indicate unstable solutions. The parameters for the chemical event chain model are given in Table~\ref{table.parameters}. The parameters for the phase model are adopted from the chemical event chain model with $\omega \approx 0.224$ estimated via Eq.~(\ref{eq.omega.est}) and $\varepsilon = r \beta$ with $r=1.33 \times 10^{-4}$.}
\label{fig.comparison}
\end{figure*}

\subsection{Comparison of phase oscillator model and chemical event chain model}
\label{subsec:comparison}
\noindent
We now show that the phase oscillator model Eq.~(\ref{stos.phaseoscillators}) can capture the key features of the collective modes described by the chemical event chain model Eq.~(\ref{eq.me}).
We compare the collective frequency~$\Omega$, Eq.~(\ref{eq.collectivefrequency}), obtained from the phase model to the frequency spectrum of oscillations from the chemical event chain model.
For the distribution~$g$ of delay times in the phase model, we adopt the parameter values of $\lambda$ and $n$ used in the chemical event chain model.
For the intrinsic frequency~$\omega$ in the phase model, we use an estimate provided in Ref.~\cite{morelli07} for a single uncoupled oscillator of the same type as investigated here,
\begin{align}
	\omega \simeq \frac{\pi}{\tilde n\tilde\lambda^{-1}+\tilde\kappa^{-1}} \ . \label{eq.omega.est}
\end{align}
 The coupling strength~$\varepsilon$ is the only parameter in the phase model whose relationship to the kinetic parameters of the event chain model is not obvious.
For simplicity, we here assume that~$\varepsilon$ scales linearly with the activation strength~$\beta$ in the chemical event chain model (see also Section~\ref{subsec:coupling}) and fix the ratio $r = \varepsilon/\beta$ by hand.

We assess whether the synchronized states of both models agree by comparing the frequency and stability solutions of the phase model to periodograms of the chemical event chain system \footnote{The logarithmic periodogram $\mathcal{P}$ of a time series $(x_1,\hdots,x_m)$ is given by $\mathcal{P}_\omega=2 \ln |\hat x_\omega|$, where $\hat x_\omega$ is the discrete Fourier transform of $x_k$.}.
Figs.~\ref{fig.comparison}a--d show such periodograms for different activation strengths $\beta$, with bright regions indicating strong frequency components.
Figs.~\ref{fig.comparison}e--h show the same density plots, superimposed with solutions for the collective frequency~$\Omega$, Eq.~(\ref{eq.collectivefrequency}).
The dominant frequency components exhibit characteristic jumps at delays being odd multiples of $T/4$, where $T$ is the uncoupled period, as already observed earlier, compare to Figs.~\ref{fig.modes}d--f.
This is expected because of stochastic switching between in-phase and anti-phase synchrony with different frequencies, see Section~\ref{subsec:switching}.
Moreover, as the signaling delay~$\tau$ increases, the dominant frequency components approach the intrinsic frequency of the uncoupled oscillator, a behavior that the phase model captures as described in the previous section and Fig.~\ref{fig.phase.oscillator.frequency}a.
Interestingly, the phase model exhibits regions in which the in-phase and anti-phase solution are simultaneously stable.
This implies that sufficiently strong fluctuations can drive the system out of one synchronized state into the basin of attraction of the other, consistent with the stochastic switching between in-phase and anti-phase synchrony found in the chemical event chain model.
To obtain a proxy for the size of the basins of attraction of the two states in the phase oscillator model, we numerically solve the deterministic Eq.~({\ref{stos.phaseoscillators}}) with a constant phase difference $\Delta\phi$ between the two oscillators as an initial history, $(\phi_1-\phi_2)|_{t\leq 0}=\Delta\phi$, and monitor their long-time phase difference to determine their final state \footnote{Note that the phase model Eq.~({\ref{stos.phaseoscillators}}) is an infinite-dimensional system due to the presence of delays and therefore requires an entire phase history $\phi_i(t)|_{t\leq 0}$ as an initial condition \mbox{\cite{schuster89}}. Therefore, the final state of the system may depend on the entire time dependence of the initial history. For simplicity, we here only focus on constant initial conditions.}. The region plots in Fig.~{\ref{fig.phase.oscillator.frequency}} display these the final states depending on $\Delta\phi$ and illustrate how the relative size of such basins change as the mean signaling delay~$\tau$ is varied.

\section{Discussion}	\noindent
Considering two coupled genetic oscillators,
we have shown how stochastic coupling by signaling chains affects their frequency and quality and promotes synchronization.
An important feature of the chemical event chain framework is that it naturally accounts for distributed signaling and transcriptional delays that are a consequence of the sequences of chemical steps.
These distributed signaling delays have profound consequences for oscillator dynamics and fluctuations that cannot be captured by simplified descriptions such as rate equations or coupled oscillators with a discrete time delay.
In particular, we found that synchrony and quality are maximized in isolated islands in parameter space characterizing coupling delay and coupling strength.
Moreover, noisy coupling can lead to stochastic switching between in-phase and anti-phase states, a behavior also found in other coupled noisy oscillators, e.g., Hodgkin--Huxley neurons~\cite{ao13} and delay-coupled phase oscillators~\cite{dhuys14}.
Our findings may shed light on the operating regime of cellular genetic oscillator systems in which precise timing is vital, such as the circadian clock \cite{schibler05} and the vertebrate segmentation clock \cite{oates12}.
The Delta--Notch signaling pathway may provide an experimental system where in-phase and anti-phase oscillations have been observed in the context of the segmentation clock and neurogenesis respectively~\cite{shimojo16, shimojo16b}. Stochastic mode switching could be explored in this system using synthetic approaches~\cite{matsuda15} and optogenetic perturbations~\cite{isomura17}.
Here we have chosen intercellular coupling as an example, but coupled genetic oscillators also occur within cells \mbox{\cite{schroter12}}, or, on a coarse-grained level, as coupled subpopulations of oscillators, such as different regions of the mammalian circadian clock \mbox{\cite{bernard07,azzi17}}, for which our modeling framework can be adapted as well. Moreover, using a unidirectional signaling mechanism, effects of stochastic signaling on entrainment to external signals can be studied, an aspect relevant for circadian clock research \mbox{\cite{abraham10,erzberger13,ananthasubramaniam14}}.

Key effects of distributed signaling delays that result from chemical chains can be captured by an effective phase oscillator model.
This phase model can also be extended to include noise which enables to study precision of collective oscillations in a simplified picture.

Our results demonstrate the interplay of stochasticity and nonlinear effects in genetic regulatory networks containing chemical event chains.
It extends existing approaches to represent fluctuations in biochemical systems and captures the statistics of non-equilibrium noise that arises in such chemical processes.
This approach is not limited to oscillatory systems investigated here but can also be applied to other genetic feedback systems such as homeostatic systems, switches, and feed-forward cascades.

\begin{acknowledgments}
\noindent We thank Andy Oates for inspiring discussions. LGM acknowledges support from ANPCyT PICT 2012 1954, PICT 2013 1301 and FOCEM Mercorsur (COF 03/11).
\end{acknowledgments}

\begin{appendix}	

\section{Stochastic simulations}%
\label{sec:simulations}
\noindent Direct numerical solutions of the master equation, Eq.~(\ref{eq.me}), is impracticable due to the high dimensionality of the state space.
Instead, a stochastic simulation algorithm of the Gillespie-type has been used to compute exact realizations of trajectories of the model~\cite{gillespie77}.
Expectation values were obtained by computing averages of the respective observable over multiple realizations.
The data shown in Figs.~\ref{fig.modes} and \ref{fig.quality} were obtained by averaging over 50 realizations of duration 80\,000 units of time for each data point.

\section{Wavelet transform}
\label{sec:wavelet.transform}

\noindent The continuous wavelet transform of a discrete time series $(x_1,\hdots,x_m)$ sampled with time interval $\delta t$ is defined by  \cite{torrence98}
\begin{align}
	W(s,k) = \frac{1}{\sqrt{s}}\sum_{j=1}^m x_j \Psi^* \bigg( \frac{j-k}{s} \bigg) \ ,
\end{align}
where $s$ is the wavelet scale.
We here choose the Gabor wavelet function, given by $\Psi(t)=\pi^{-1/4}\smash{\mathrm{e}^{6\mathrm{i}t-t^2/2}}$. For the Gabor wavelet, the scale $s$ corresponds to a period of $T(s) \approx 1.033 s \cdot \delta t$.
The wavelet scalogram in Fig.~\ref{fig.switching}c displays the squared magnitude $|W(s,k)|^2$ as a density plot, where the abscissa shows time~$t=k \cdot \delta t$ and the ordinate shows  the period $T=T(s)$.

\section{Stability analysis of the synchronized state in the phase model}
\label{sec:linear.stability}
\noindent To assess the stability of the in-phase synchronized state $\phi_\mu(t)= \Omega t$, we linearize the dynamics around this state~\cite{strogatz}. We use the standard ansatz
$\phi_\mu(t) = \Omega t + \xi_\mu(t)$ in Eq.~(\ref{stos.phaseoscillators}), where $\xi_\mu$ is a small perturbation. We obtain the time evolution of the perturbation by expanding Eq.~(\ref{stos.phaseoscillators}) to first order in $\xi$, which yields
\begin{align}
	\frac{1}{\varepsilon}\frac{\mathrm{d}\xi_\mu}{\mathrm{d}t} = \int_0^\infty g(s) \cos(\Omega s) (\xi_{\bar\mu}^{(s)} - \xi_\mu) \, \mathrm{d}s \ ,
\end{align}
where $\bar \mu$ refers to the respective other oscillator as in Eq.~(\ref{eq.me}) and where we have defined the delayed variable $\xi_{\mu}^{(s)}(t) = \xi_\mu(t-s)$.
We decouple the dynamics by defining the collective modes~$\varphi_k=\xi_1 + k \xi_2$ with $k=+1,-1$.
Inverting this definition yields $\xi_1 = (\varphi_+ + \varphi_-)/2$ and $\xi_2 = (\varphi_+ - \varphi_-)/2$, which shows that exciting the collective mode~$\varphi_+$ shifts both oscillators by the same amount and thus corresponds to a global phase shift, whereas~$\varphi_-$ is the phase difference between both oscillators.
The dynamics of these collective modes are given by
\begin{align}
	\frac{1}{\varepsilon}\frac{\mathrm{d}\varphi_k}{\mathrm{d}t} = \int_0^\infty g(s) \cos(\Omega s) ( k\varphi_{k}^{(s)} - \varphi_k ) \, \mathrm{d}s \ . \label{stos.collectivemodes.dynamics}
\end{align}
The characteristic equation for these modes is obtained using the exponential ansatz $\varphi_k(t)=\mathrm{e}^{\eta_k t}$ with $\eta_k$ being complex. The sign of $\operatorname{Re}\eta_k$ then determines whether perturbations decay ($\operatorname{Re}\eta_k<0$) or grow ($\operatorname{Re}\eta_k>0$) and thus whether the synchronized state is stable or unstable~\cite{amann07}. Using this ansatz in Eq.~(\ref{stos.collectivemodes.dynamics}), we obtain
\begin{align}
	\frac{\eta_k}{\varepsilon}  = \int_0^\infty g(s) \cos(\Omega s) (k\mathrm{e}^{-\eta_k s}-1 ) \, \mathrm{d}s \ .\label{stos.phaseoscillators.ceq}
\end{align}
Since the Gamma distribution~$g$, Eq.~(\ref{eq.gamma}), decays as $\mathrm{e}^{-\lambda s}$, the integral on the rhs of Eq.~(\ref{stos.phaseoscillators.ceq}) is only well-defined if $\operatorname{Re} \eta_k > - \lambda$.
In this case, the integral can be solved analytically and the resulting characteristic equation is
\begin{align}
	k \Gamma(\eta_k)  - {\eta_k}/\varepsilon =  \Gamma(0) \ , \label{stos.phaseoscillators.ceq.sol}
\end{align}
where
\begin{align}
	\Gamma(\eta)	= \frac{\lambda^n}{2} \left[ \frac{1}{(\lambda+\eta+\mathrm{i}\Omega)^n} + \frac{1}{(\lambda+\eta-\mathrm{i}\Omega)^n} \right] \ .
\end{align}
In general, Eq.~(\ref{stos.phaseoscillators.ceq.sol}) can have multiple solutions in $\eta_k$.
The synchronized state is linearly stable if and only if $\operatorname{Re}\eta_k<0$ holds for all solutions~$\eta_k$ to Eq.~(\ref{stos.phaseoscillators.ceq}) for both $k=+1$ and $k=-1$.
The stability of the anti-phase synchronized state is determined by Eq.~(\ref{stos.phaseoscillators.ceq.sol}) with $\varepsilon$ replaced by $-\varepsilon$.
For Figs.~\ref{fig.phase.oscillator.frequency} and \ref{fig.comparison}, we determine the stability of a given synchronized state by solving Eq.~(\ref{stos.phaseoscillators.ceq.sol}) numerically and determining the sign of the solution with largest real part.

\end{appendix}


\begin{thebibliography}{71}%
\makeatletter
\providecommand \@ifxundefined [1]{%
 \@ifx{#1\undefined}
}%
\providecommand \@ifnum [1]{%
 \ifnum #1\expandafter \@firstoftwo
 \else \expandafter \@secondoftwo
 \fi
}%
\providecommand \@ifx [1]{%
 \ifx #1\expandafter \@firstoftwo
 \else \expandafter \@secondoftwo
 \fi
}%
\providecommand \natexlab [1]{#1}%
\providecommand \enquote  [1]{``#1''}%
\providecommand \bibnamefont  [1]{#1}%
\providecommand \bibfnamefont [1]{#1}%
\providecommand \citenamefont [1]{#1}%
\providecommand \href@noop [0]{\@secondoftwo}%
\providecommand \href [0]{\begingroup \@sanitize@url \@href}%
\providecommand \@href[1]{\@@startlink{#1}\@@href}%
\providecommand \@@href[1]{\endgroup#1\@@endlink}%
\providecommand \@sanitize@url [0]{\catcode `\\12\catcode `\$12\catcode
  `\&12\catcode `\#12\catcode `\^12\catcode `\_12\catcode `\%12\relax}%
\providecommand \@@startlink[1]{}%
\providecommand \@@endlink[0]{}%
\providecommand \url  [0]{\begingroup\@sanitize@url \@url }%
\providecommand \@url [1]{\endgroup\@href {#1}{\urlprefix }}%
\providecommand \urlprefix  [0]{URL }%
\providecommand \Eprint [0]{\href }%
\providecommand \doibase [0]{http://dx.doi.org/}%
\providecommand \selectlanguage [0]{\@gobble}%
\providecommand \bibinfo  [0]{\@secondoftwo}%
\providecommand \bibfield  [0]{\@secondoftwo}%
\providecommand \translation [1]{[#1]}%
\providecommand \BibitemOpen [0]{}%
\providecommand \bibitemStop [0]{}%
\providecommand \bibitemNoStop [0]{.\EOS\space}%
\providecommand \EOS [0]{\spacefactor3000\relax}%
\providecommand \BibitemShut  [1]{\csname bibitem#1\endcsname}%
\let\auto@bib@innerbib\@empty
%</preamble>
\bibitem [{\citenamefont {Alon}(2006)}]{alon}%
  \BibitemOpen
  \bibfield  {author} {\bibinfo {author} {\bibfnamefont {U.}~\bibnamefont
  {Alon}},\ }\href@noop {} {\emph {\bibinfo {title} {An Introduction to Systems
  Biology: Design Principles of Biological Circuits}}}\ (\bibinfo  {publisher}
  {Chapman \& Hall/CRC Press},\ \bibinfo {address} {Boca Raton, Florida},\
  \bibinfo {year} {2006})\BibitemShut {NoStop}%
\bibitem [{\citenamefont {Goldbeter}(1997)}]{goldbeter}%
  \BibitemOpen
  \bibfield  {author} {\bibinfo {author} {\bibfnamefont {A.}~\bibnamefont
  {Goldbeter}},\ }\href@noop {} {\emph {\bibinfo {title} {Biochemical
  Oscillations and Cellular Rhythms: The Molecular Bases of Periodic and
  Chaotic Behaviour}}}\ (\bibinfo  {publisher} {Cambridge University Press},\
  \bibinfo {year} {1997})\BibitemShut {NoStop}%
\bibitem [{\citenamefont {Alberts}\ \emph {et~al.}(2002)\citenamefont
  {Alberts}, \citenamefont {Johnson}, \citenamefont {Lewis}, \citenamefont
  {Raff}, \citenamefont {Roberts},\ and\ \citenamefont {Walter}}]{alberts}%
  \BibitemOpen
  \bibfield  {author} {\bibinfo {author} {\bibfnamefont {B.}~\bibnamefont
  {Alberts}}, \bibinfo {author} {\bibfnamefont {A.}~\bibnamefont {Johnson}},
  \bibinfo {author} {\bibfnamefont {J.}~\bibnamefont {Lewis}}, \bibinfo
  {author} {\bibfnamefont {M.}~\bibnamefont {Raff}}, \bibinfo {author}
  {\bibfnamefont {K.}~\bibnamefont {Roberts}}, \ and\ \bibinfo {author}
  {\bibfnamefont {P.}~\bibnamefont {Walter}},\ }\href@noop {} {\emph {\bibinfo
  {title} {Molecular Biology of the Cell}}},\ \bibinfo {edition} {4th}\ ed.\
  (\bibinfo  {publisher} {Garland Science},\ \bibinfo {address} {New York},\
  \bibinfo {year} {2002})\BibitemShut {NoStop}%
\bibitem [{\citenamefont {Ilagan}\ and\ \citenamefont
  {Kopan}(2007)}]{ilagan07}%
  \BibitemOpen
  \bibfield  {author} {\bibinfo {author} {\bibfnamefont {M.~X.~G.}\
  \bibnamefont {Ilagan}}\ and\ \bibinfo {author} {\bibfnamefont
  {R.}~\bibnamefont {Kopan}},\ }\bibfield  {title} {\enquote {\bibinfo {title}
  {{SnapShot: Notch signaling pathway}},}\ }\href@noop {} {\bibfield  {journal}
  {\bibinfo  {journal} {Cell}\ }\textbf {\bibinfo {volume} {128}},\ \bibinfo
  {pages} {1246} (\bibinfo {year} {2007})}\BibitemShut {NoStop}%
\bibitem [{\citenamefont {Nov{\'{a}}k}\ and\ \citenamefont
  {Tyson}(2008)}]{novak08}%
  \BibitemOpen
  \bibfield  {author} {\bibinfo {author} {\bibfnamefont {B.}~\bibnamefont
  {Nov{\'{a}}k}}\ and\ \bibinfo {author} {\bibfnamefont {J.~J.}\ \bibnamefont
  {Tyson}},\ }\bibfield  {title} {\enquote {\bibinfo {title} {Design principles
  of biochemical oscillators},}\ }\href@noop {} {\bibfield  {journal} {\bibinfo
   {journal} {Nat. Rev. Mol. Cell Biol.}\ }\textbf {\bibinfo {volume} {9}},\
  \bibinfo {pages} {981--991} (\bibinfo {year} {2008})}\BibitemShut {NoStop}%
\bibitem [{\citenamefont {Morelli}\ and\ \citenamefont
  {J{\"{u}}licher}(2007)}]{morelli07}%
  \BibitemOpen
  \bibfield  {author} {\bibinfo {author} {\bibfnamefont {L.~G.}\ \bibnamefont
  {Morelli}}\ and\ \bibinfo {author} {\bibfnamefont {F.}~\bibnamefont
  {J{\"{u}}licher}},\ }\bibfield  {title} {\enquote {\bibinfo {title}
  {Precision of genetic oscillators and clocks},}\ }\href@noop {} {\bibfield
  {journal} {\bibinfo  {journal} {Phys. Rev. Lett.}\ }\textbf {\bibinfo
  {volume} {98}},\ \bibinfo {pages} {228101} (\bibinfo {year}
  {2007})}\BibitemShut {NoStop}%
\bibitem [{\citenamefont {Hardin}\ \emph {et~al.}(1990)\citenamefont {Hardin},
  \citenamefont {Hall},\ and\ \citenamefont {Rosbash}}]{hardin90}%
  \BibitemOpen
  \bibfield  {author} {\bibinfo {author} {\bibfnamefont {P.~E.}\ \bibnamefont
  {Hardin}}, \bibinfo {author} {\bibfnamefont {J.~C.}\ \bibnamefont {Hall}}, \
  and\ \bibinfo {author} {\bibfnamefont {M.}~\bibnamefont {Rosbash}},\
  }\bibfield  {title} {\enquote {\bibinfo {title} {{Feedback of the Drosophila
  period gene product on circadian cycling of its messenger RNA levels.}}}\
  }\href@noop {} {\bibfield  {journal} {\bibinfo  {journal} {Nature}\ }\textbf
  {\bibinfo {volume} {343}},\ \bibinfo {pages} {536--540} (\bibinfo {year}
  {1990})}\BibitemShut {NoStop}%
\bibitem [{\citenamefont {Dunlap}(1999)}]{dunlap99}%
  \BibitemOpen
  \bibfield  {author} {\bibinfo {author} {\bibfnamefont {J.~C.}\ \bibnamefont
  {Dunlap}},\ }\bibfield  {title} {\enquote {\bibinfo {title} {{Molecular bases
  for circadian clocks}},}\ }\href@noop {} {\bibfield  {journal} {\bibinfo
  {journal} {Cell}\ }\textbf {\bibinfo {volume} {96}},\ \bibinfo {pages}
  {271--290} (\bibinfo {year} {1999})}\BibitemShut {NoStop}%
\bibitem [{\citenamefont {Smolen}\ \emph {et~al.}(2002)\citenamefont {Smolen},
  \citenamefont {Baxter},\ and\ \citenamefont {Byrne}}]{smolen02}%
  \BibitemOpen
  \bibfield  {author} {\bibinfo {author} {\bibfnamefont {P.}~\bibnamefont
  {Smolen}}, \bibinfo {author} {\bibfnamefont {D.~A.}\ \bibnamefont {Baxter}},
  \ and\ \bibinfo {author} {\bibfnamefont {J.~H.}\ \bibnamefont {Byrne}},\
  }\bibfield  {title} {\enquote {\bibinfo {title} {{A Reduced Model Clarifies
  the Role of Feedback Loops and Time Delays in the Drosophila Circadian
  Oscillator}},}\ }\href@noop {} {\bibfield  {journal} {\bibinfo  {journal}
  {Biophys. J.}\ }\textbf {\bibinfo {volume} {83}},\ \bibinfo {pages}
  {2349--2359} (\bibinfo {year} {2002})}\BibitemShut {NoStop}%
\bibitem [{\citenamefont {Schibler}\ and\ \citenamefont
  {Naef}(2005)}]{schibler05}%
  \BibitemOpen
  \bibfield  {author} {\bibinfo {author} {\bibfnamefont {U.}~\bibnamefont
  {Schibler}}\ and\ \bibinfo {author} {\bibfnamefont {F.}~\bibnamefont
  {Naef}},\ }\bibfield  {title} {\enquote {\bibinfo {title} {{Cellular
  oscillators: Rhythmic gene expression and metabolism}},}\ }\href@noop {}
  {\bibfield  {journal} {\bibinfo  {journal} {Curr. Opin. Cell Biol.}\ }\textbf
  {\bibinfo {volume} {17}},\ \bibinfo {pages} {223--229} (\bibinfo {year}
  {2005})}\BibitemShut {NoStop}%
\bibitem [{\citenamefont {Hastings}\ \emph {et~al.}(2007)\citenamefont
  {Hastings}, \citenamefont {O'Neill},\ and\ \citenamefont
  {Maywood}}]{hastings07}%
  \BibitemOpen
  \bibfield  {author} {\bibinfo {author} {\bibfnamefont {M.}~\bibnamefont
  {Hastings}}, \bibinfo {author} {\bibfnamefont {J.~S.}\ \bibnamefont
  {O'Neill}}, \ and\ \bibinfo {author} {\bibfnamefont {E.~S.}\ \bibnamefont
  {Maywood}},\ }\bibfield  {title} {\enquote {\bibinfo {title} {{Circadian
  clocks: regulators of endocrine and metabolic rhythms}},}\ }\href@noop {}
  {\bibfield  {journal} {\bibinfo  {journal} {J. Endocrinol.}\ }\textbf
  {\bibinfo {volume} {195}},\ \bibinfo {pages} {187--198} (\bibinfo {year}
  {2007})}\BibitemShut {NoStop}%
\bibitem [{\citenamefont {Zwicker}\ \emph {et~al.}(2010)\citenamefont
  {Zwicker}, \citenamefont {Lubensky},\ and\ \citenamefont {ten
  Wolde}}]{zwicker10}%
  \BibitemOpen
  \bibfield  {author} {\bibinfo {author} {\bibfnamefont {D.}~\bibnamefont
  {Zwicker}}, \bibinfo {author} {\bibfnamefont {D.~K.}\ \bibnamefont
  {Lubensky}}, \ and\ \bibinfo {author} {\bibfnamefont {P.~R.}\ \bibnamefont
  {ten Wolde}},\ }\bibfield  {title} {\enquote {\bibinfo {title} {{Robust
  circadian clocks from coupled protein-modification and
  transcription--translation cycles}},}\ }\href@noop {} {\bibfield  {journal}
  {\bibinfo  {journal} {Proc. Natl. Acad. Sci. USA}\ }\textbf {\bibinfo
  {volume} {107}} (\bibinfo {year} {2010})}\BibitemShut {NoStop}%
\bibitem [{\citenamefont {Zhang}\ and\ \citenamefont {Kay}(2010)}]{zhang10b}%
  \BibitemOpen
  \bibfield  {author} {\bibinfo {author} {\bibfnamefont {E.~E.}\ \bibnamefont
  {Zhang}}\ and\ \bibinfo {author} {\bibfnamefont {S.~A.}\ \bibnamefont
  {Kay}},\ }\bibfield  {title} {\enquote {\bibinfo {title} {Clocks not winding
  down: unravelling circadian networks},}\ }\href@noop {} {\bibfield  {journal}
  {\bibinfo  {journal} {Nat. Rev. Mol. Cell Biol.}\ }\textbf {\bibinfo {volume}
  {11}},\ \bibinfo {pages} {764--776} (\bibinfo {year} {2010})}\BibitemShut
  {NoStop}%
\bibitem [{\citenamefont {Abraham}\ \emph {et~al.}(2010)\citenamefont
  {Abraham}, \citenamefont {Granada}, \citenamefont {Westermark}, \citenamefont
  {Heine}, \citenamefont {Herzel},\ and\ \citenamefont {Kramer}}]{abraham10}%
  \BibitemOpen
  \bibfield  {author} {\bibinfo {author} {\bibfnamefont {U.}~\bibnamefont
  {Abraham}}, \bibinfo {author} {\bibfnamefont {A.~E.}\ \bibnamefont
  {Granada}}, \bibinfo {author} {\bibfnamefont {P.~O.}\ \bibnamefont
  {Westermark}}, \bibinfo {author} {\bibfnamefont {M.}~\bibnamefont {Heine}},
  \bibinfo {author} {\bibfnamefont {H.}~\bibnamefont {Herzel}}, \ and\ \bibinfo
  {author} {\bibfnamefont {A.}~\bibnamefont {Kramer}},\ }\bibfield  {title}
  {\enquote {\bibinfo {title} {{Coupling governs entrainment range of circadian
  clocks}},}\ }\href@noop {} {\bibfield  {journal} {\bibinfo  {journal} {Mol.
  Sys. Biol.}\ }\textbf {\bibinfo {volume} {6}},\ \bibinfo {pages} {438}
  (\bibinfo {year} {2010})}\BibitemShut {NoStop}%
\bibitem [{\citenamefont {Granada}\ \emph {et~al.}(2013)\citenamefont
  {Granada}, \citenamefont {Bordyugov}, \citenamefont {Kramer},\ and\
  \citenamefont {Herzel}}]{granada13}%
  \BibitemOpen
  \bibfield  {author} {\bibinfo {author} {\bibfnamefont {A.~E.}\ \bibnamefont
  {Granada}}, \bibinfo {author} {\bibfnamefont {G.}~\bibnamefont {Bordyugov}},
  \bibinfo {author} {\bibfnamefont {A.}~\bibnamefont {Kramer}}, \ and\ \bibinfo
  {author} {\bibfnamefont {H.}~\bibnamefont {Herzel}},\ }\bibfield  {title}
  {\enquote {\bibinfo {title} {{Human Chronotypes from a Theoretical
  Perspective}},}\ }\href@noop {} {\bibfield  {journal} {\bibinfo  {journal}
  {PLOS ONE}\ }\textbf {\bibinfo {volume} {8}},\ \bibinfo {pages} {e59464}
  (\bibinfo {year} {2013})}\BibitemShut {NoStop}%
\bibitem [{\citenamefont {Imayoshi}\ and\ \citenamefont
  {Kageyama}(2014)}]{imayoshi14}%
  \BibitemOpen
  \bibfield  {author} {\bibinfo {author} {\bibfnamefont {I.}~\bibnamefont
  {Imayoshi}}\ and\ \bibinfo {author} {\bibfnamefont {R.}~\bibnamefont
  {Kageyama}},\ }\bibfield  {title} {\enquote {\bibinfo {title} {{bHLH Factors
  in Self-Renewal, Multipotency, and Fate Choice of Neural Progenitor
  Cells}},}\ }\href@noop {} {\bibfield  {journal} {\bibinfo  {journal}
  {Neuron}\ }\textbf {\bibinfo {volume} {82}},\ \bibinfo {pages} {9--23}
  (\bibinfo {year} {2014})}\BibitemShut {NoStop}%
\bibitem [{\citenamefont {Shimojo}\ and\ \citenamefont
  {Kageyama}(2016)}]{shimojo16b}%
  \BibitemOpen
  \bibfield  {author} {\bibinfo {author} {\bibfnamefont {H.}~\bibnamefont
  {Shimojo}}\ and\ \bibinfo {author} {\bibfnamefont {R.}~\bibnamefont
  {Kageyama}},\ }\bibfield  {title} {\enquote {\bibinfo {title} {{Oscillatory
  control of Delta-like1 in somitogenesis and neurogenesis: A unified model for
  different oscillatory dynamics}},}\ }\href@noop {} {\bibfield  {journal}
  {\bibinfo  {journal} {Semin. Cell Dev. Biol.}\ }\textbf {\bibinfo {volume}
  {49}},\ \bibinfo {pages} {76--82} (\bibinfo {year} {2016})}\BibitemShut
  {NoStop}%
\bibitem [{\citenamefont {J\"org}\ \emph {et~al.}(2015)\citenamefont {J\"org},
  \citenamefont {Morelli}, \citenamefont {Soroldoni}, \citenamefont {Oates},\
  and\ \citenamefont {J\"ulicher}}]{jorg15}%
  \BibitemOpen
  \bibfield  {author} {\bibinfo {author} {\bibfnamefont {D.~J.}\ \bibnamefont
  {J\"org}}, \bibinfo {author} {\bibfnamefont {L.~G.}\ \bibnamefont {Morelli}},
  \bibinfo {author} {\bibfnamefont {D.}~\bibnamefont {Soroldoni}}, \bibinfo
  {author} {\bibfnamefont {A.~C.}\ \bibnamefont {Oates}}, \ and\ \bibinfo
  {author} {\bibfnamefont {F.}~\bibnamefont {J\"ulicher}},\ }\bibfield  {title}
  {\enquote {\bibinfo {title} {{Continuum theory of gene expression waves
  during vertebrate segmentation}},}\ }\href@noop {} {\bibfield  {journal}
  {\bibinfo  {journal} {New J. Phys.}\ }\textbf {\bibinfo {volume} {17}},\
  \bibinfo {pages} {093042} (\bibinfo {year} {2015})}\BibitemShut {NoStop}%
\bibitem [{\citenamefont {Oates}\ \emph {et~al.}(2012)\citenamefont {Oates},
  \citenamefont {Morelli},\ and\ \citenamefont {Ares}}]{oates12}%
  \BibitemOpen
  \bibfield  {author} {\bibinfo {author} {\bibfnamefont {A.~C.}\ \bibnamefont
  {Oates}}, \bibinfo {author} {\bibfnamefont {L.~G.}\ \bibnamefont {Morelli}},
  \ and\ \bibinfo {author} {\bibfnamefont {S.}~\bibnamefont {Ares}},\
  }\bibfield  {title} {\enquote {\bibinfo {title} {Patterning embryos with
  oscillations: structure, function and dynamics of the vertebrate segmentation
  clock},}\ }\href@noop {} {\bibfield  {journal} {\bibinfo  {journal}
  {Development}\ }\textbf {\bibinfo {volume} {139}},\ \bibinfo {pages}
  {625--639} (\bibinfo {year} {2012})}\BibitemShut {NoStop}%
\bibitem [{\citenamefont {Niederholtmeyer}\ \emph {et~al.}(2015)\citenamefont
  {Niederholtmeyer}, \citenamefont {Sun}, \citenamefont {Hori}, \citenamefont
  {Yeung}, \citenamefont {Verpoorte}, \citenamefont {Murray},\ and\
  \citenamefont {Maerkl}}]{niederholtmeyer15}%
  \BibitemOpen
  \bibfield  {author} {\bibinfo {author} {\bibfnamefont {H.}~\bibnamefont
  {Niederholtmeyer}}, \bibinfo {author} {\bibfnamefont {Z.~Z.}\ \bibnamefont
  {Sun}}, \bibinfo {author} {\bibfnamefont {Y.}~\bibnamefont {Hori}}, \bibinfo
  {author} {\bibfnamefont {E.}~\bibnamefont {Yeung}}, \bibinfo {author}
  {\bibfnamefont {A.}~\bibnamefont {Verpoorte}}, \bibinfo {author}
  {\bibfnamefont {R.~M.}\ \bibnamefont {Murray}}, \ and\ \bibinfo {author}
  {\bibfnamefont {S.~J.}\ \bibnamefont {Maerkl}},\ }\bibfield  {title}
  {\enquote {\bibinfo {title} {{Rapid cell-free forward engineering of novel
  genetic ring oscillators}},}\ }\href@noop {} {\bibfield  {journal} {\bibinfo
  {journal} {eLife}\ }\textbf {\bibinfo {volume} {4}},\ \bibinfo {pages}
  {e09771} (\bibinfo {year} {2015})}\BibitemShut {NoStop}%
\bibitem [{\citenamefont {Tokuda}\ \emph {et~al.}(2015)\citenamefont {Tokuda},
  \citenamefont {Ono}, \citenamefont {Ananthasubramaniam}, \citenamefont
  {Honma}, \citenamefont {Honma},\ and\ \citenamefont {Herzel}}]{tokuda15}%
  \BibitemOpen
  \bibfield  {author} {\bibinfo {author} {\bibfnamefont {I.~T.}\ \bibnamefont
  {Tokuda}}, \bibinfo {author} {\bibfnamefont {D.}~\bibnamefont {Ono}},
  \bibinfo {author} {\bibfnamefont {B.}~\bibnamefont {Ananthasubramaniam}},
  \bibinfo {author} {\bibfnamefont {S.}~\bibnamefont {Honma}}, \bibinfo
  {author} {\bibfnamefont {K.-I.}\ \bibnamefont {Honma}}, \ and\ \bibinfo
  {author} {\bibfnamefont {H.}~\bibnamefont {Herzel}},\ }\bibfield  {title}
  {\enquote {\bibinfo {title} {{Coupling Controls the Synchrony of Clock Cells
  in Development and Knockouts}},}\ }\href@noop {} {\bibfield  {journal}
  {\bibinfo  {journal} {Biophysical Journal}\ }\textbf {\bibinfo {volume}
  {109}},\ \bibinfo {pages} {2159} (\bibinfo {year} {2015})}\BibitemShut
  {NoStop}%
\bibitem [{\citenamefont {J{\"o}rg}(2017)}]{jorg17b}%
  \BibitemOpen
  \bibfield  {author} {\bibinfo {author} {\bibfnamefont {D.~J.}\ \bibnamefont
  {J{\"o}rg}},\ }\bibfield  {title} {\enquote {\bibinfo {title} {{Amplitude
  bounds for biochemical oscillators}},}\ }\href@noop {} {\bibfield  {journal}
  {\bibinfo  {journal} {EPL (Europhys. Lett.)}\ }\textbf {\bibinfo {volume}
  {119}},\ \bibinfo {pages} {58004} (\bibinfo {year} {2017})}\BibitemShut
  {NoStop}%
\bibitem [{\citenamefont {Elowitz}\ and\ \citenamefont
  {Leibler}(2000)}]{elowitz00}%
  \BibitemOpen
  \bibfield  {author} {\bibinfo {author} {\bibfnamefont {M.~B.}\ \bibnamefont
  {Elowitz}}\ and\ \bibinfo {author} {\bibfnamefont {S.}~\bibnamefont
  {Leibler}},\ }\bibfield  {title} {\enquote {\bibinfo {title} {{A synthetic
  oscillatory network of transcriptional regulators}},}\ }\href@noop {}
  {\bibfield  {journal} {\bibinfo  {journal} {Nature}\ }\textbf {\bibinfo
  {volume} {403}},\ \bibinfo {pages} {335--338} (\bibinfo {year}
  {2000})}\BibitemShut {NoStop}%
\bibitem [{\citenamefont {Garcia-Ojalvo}\ \emph {et~al.}(2004)\citenamefont
  {Garcia-Ojalvo}, \citenamefont {Elowitz},\ and\ \citenamefont
  {Strogatz}}]{garciaojalvo04}%
  \BibitemOpen
  \bibfield  {author} {\bibinfo {author} {\bibfnamefont {J.}~\bibnamefont
  {Garcia-Ojalvo}}, \bibinfo {author} {\bibfnamefont {M.~B.}\ \bibnamefont
  {Elowitz}}, \ and\ \bibinfo {author} {\bibfnamefont {S.~H.}\ \bibnamefont
  {Strogatz}},\ }\bibfield  {title} {\enquote {\bibinfo {title} {{Modeling a
  synthetic multicellular clock: repressilators coupled by quorum sensing.}}}\
  }\href@noop {} {\bibfield  {journal} {\bibinfo  {journal} {Proc. Natl. Acad.
  Sci. USA}\ }\textbf {\bibinfo {volume} {101}},\ \bibinfo {pages} {10955}
  (\bibinfo {year} {2004})}\BibitemShut {NoStop}%
\bibitem [{\citenamefont {Stricker}\ \emph {et~al.}(2008)\citenamefont
  {Stricker}, \citenamefont {Cookson}, \citenamefont {Bennett}, \citenamefont
  {Mather}, \citenamefont {Tsimring},\ and\ \citenamefont
  {Hasty}}]{stricker08}%
  \BibitemOpen
  \bibfield  {author} {\bibinfo {author} {\bibfnamefont {J.}~\bibnamefont
  {Stricker}}, \bibinfo {author} {\bibfnamefont {S.}~\bibnamefont {Cookson}},
  \bibinfo {author} {\bibfnamefont {M.~R.}\ \bibnamefont {Bennett}}, \bibinfo
  {author} {\bibfnamefont {W.~H.}\ \bibnamefont {Mather}}, \bibinfo {author}
  {\bibfnamefont {L.~S.}\ \bibnamefont {Tsimring}}, \ and\ \bibinfo {author}
  {\bibfnamefont {J.}~\bibnamefont {Hasty}},\ }\bibfield  {title} {\enquote
  {\bibinfo {title} {{A fast, robust and tunable synthetic gene oscillator}},}\
  }\href@noop {} {\bibfield  {journal} {\bibinfo  {journal} {Nature}\ }\textbf
  {\bibinfo {volume} {456}},\ \bibinfo {pages} {516--519} (\bibinfo {year}
  {2008})}\BibitemShut {NoStop}%
\bibitem [{\citenamefont {Danino}\ \emph {et~al.}(2010)\citenamefont {Danino},
  \citenamefont {Mondrag\'{o}n-Palomino}, \citenamefont {Tsimring},\ and\
  \citenamefont {Hasty}}]{danino10}%
  \BibitemOpen
  \bibfield  {author} {\bibinfo {author} {\bibfnamefont {T.}~\bibnamefont
  {Danino}}, \bibinfo {author} {\bibfnamefont {O.}~\bibnamefont
  {Mondrag\'{o}n-Palomino}}, \bibinfo {author} {\bibfnamefont {L.}~\bibnamefont
  {Tsimring}}, \ and\ \bibinfo {author} {\bibfnamefont {J.}~\bibnamefont
  {Hasty}},\ }\bibfield  {title} {\enquote {\bibinfo {title} {{A synchronized
  quorum of genetic clocks}},}\ }\href@noop {} {\bibfield  {journal} {\bibinfo
  {journal} {Nature}\ }\textbf {\bibinfo {volume} {463}},\ \bibinfo {pages}
  {326--330} (\bibinfo {year} {2010})}\BibitemShut {NoStop}%
\bibitem [{\citenamefont {Kim}\ and\ \citenamefont {Winfree}(2011)}]{kim11}%
  \BibitemOpen
  \bibfield  {author} {\bibinfo {author} {\bibfnamefont {J.}~\bibnamefont
  {Kim}}\ and\ \bibinfo {author} {\bibfnamefont {E.}~\bibnamefont {Winfree}},\
  }\bibfield  {title} {\enquote {\bibinfo {title} {{Synthetic in vitro
  transcriptional oscillators}},}\ }\href@noop {} {\bibfield  {journal}
  {\bibinfo  {journal} {Mol. Sys. Biol.}\ }\textbf {\bibinfo {volume} {7}},\
  \bibinfo {pages} {465--465} (\bibinfo {year} {2011})}\BibitemShut {NoStop}%
\bibitem [{\citenamefont {Potvin-Trottier}\ \emph {et~al.}(2016)\citenamefont
  {Potvin-Trottier}, \citenamefont {Lord}, \citenamefont {Vinnicombe},\ and\
  \citenamefont {Paulsson}}]{potvintrottier16}%
  \BibitemOpen
  \bibfield  {author} {\bibinfo {author} {\bibfnamefont {L.}~\bibnamefont
  {Potvin-Trottier}}, \bibinfo {author} {\bibfnamefont {N.~D.}\ \bibnamefont
  {Lord}}, \bibinfo {author} {\bibfnamefont {G.}~\bibnamefont {Vinnicombe}}, \
  and\ \bibinfo {author} {\bibfnamefont {J.}~\bibnamefont {Paulsson}},\
  }\bibfield  {title} {\enquote {\bibinfo {title} {{Synchronous long-term
  oscillations in a synthetic gene circuit.}}}\ }\href@noop {} {\bibfield
  {journal} {\bibinfo  {journal} {Nature}\ }\textbf {\bibinfo {volume} {538}},\
  \bibinfo {pages} {514} (\bibinfo {year} {2016})}\BibitemShut {NoStop}%
\bibitem [{\citenamefont {Tayar}\ \emph {et~al.}(2017)\citenamefont {Tayar},
  \citenamefont {Karzbrun}, \citenamefont {Noireaux},\ and\ \citenamefont
  {Bar-Ziv}}]{tayar17}%
  \BibitemOpen
  \bibfield  {author} {\bibinfo {author} {\bibfnamefont {A.~M.}\ \bibnamefont
  {Tayar}}, \bibinfo {author} {\bibfnamefont {E.}~\bibnamefont {Karzbrun}},
  \bibinfo {author} {\bibfnamefont {V.}~\bibnamefont {Noireaux}}, \ and\
  \bibinfo {author} {\bibfnamefont {R.~H.}\ \bibnamefont {Bar-Ziv}},\
  }\bibfield  {title} {\enquote {\bibinfo {title} {{Synchrony and pattern
  formation of coupled genetic oscillators on a chip of artificial cells.}}}\
  }\href@noop {} {\bibfield  {journal} {\bibinfo  {journal} {Proc. Natl. Acad.
  Sci. USA}\ }\textbf {\bibinfo {volume} {114}},\ \bibinfo {pages}
  {11609--11614} (\bibinfo {year} {2017})}\BibitemShut {NoStop}%
\bibitem [{\citenamefont {Barkai}\ and\ \citenamefont
  {Leibler}(2000)}]{barkai00}%
  \BibitemOpen
  \bibfield  {author} {\bibinfo {author} {\bibfnamefont {N.}~\bibnamefont
  {Barkai}}\ and\ \bibinfo {author} {\bibfnamefont {S.}~\bibnamefont
  {Leibler}},\ }\bibfield  {title} {\enquote {\bibinfo {title} {Circadian
  clocks limited by noise},}\ }\href@noop {} {\bibfield  {journal} {\bibinfo
  {journal} {Nature}\ }\textbf {\bibinfo {volume} {403}},\ \bibinfo {pages}
  {267--268} (\bibinfo {year} {2000})}\BibitemShut {NoStop}%
\bibitem [{\citenamefont {Vilar}\ \emph {et~al.}(2002)\citenamefont {Vilar},
  \citenamefont {Kueh}, \citenamefont {Barkai},\ and\ \citenamefont
  {Leibler}}]{vilar02}%
  \BibitemOpen
  \bibfield  {author} {\bibinfo {author} {\bibfnamefont {J.~M.~G.}\
  \bibnamefont {Vilar}}, \bibinfo {author} {\bibfnamefont {H.~Y.}\ \bibnamefont
  {Kueh}}, \bibinfo {author} {\bibfnamefont {N.}~\bibnamefont {Barkai}}, \ and\
  \bibinfo {author} {\bibfnamefont {S.}~\bibnamefont {Leibler}},\ }\bibfield
  {title} {\enquote {\bibinfo {title} {Mechanisms of noise-resistance in
  genetic oscillators},}\ }\href@noop {} {\bibfield  {journal} {\bibinfo
  {journal} {Proc. Natl. Acad. Sci. USA}\ }\textbf {\bibinfo {volume} {99}},\
  \bibinfo {pages} {5988--5992} (\bibinfo {year} {2002})}\BibitemShut {NoStop}%
\bibitem [{\citenamefont {Gonze}\ \emph {et~al.}(2002)\citenamefont {Gonze},
  \citenamefont {Halloy},\ and\ \citenamefont {Gaspard}}]{gonze02}%
  \BibitemOpen
  \bibfield  {author} {\bibinfo {author} {\bibfnamefont {D.}~\bibnamefont
  {Gonze}}, \bibinfo {author} {\bibfnamefont {J.}~\bibnamefont {Halloy}}, \
  and\ \bibinfo {author} {\bibfnamefont {P.}~\bibnamefont {Gaspard}},\
  }\bibfield  {title} {\enquote {\bibinfo {title} {Biochemical clocks and
  molecular noise: Theoretical study of robustness factors},}\ }\href@noop {}
  {\bibfield  {journal} {\bibinfo  {journal} {J. Chem. Phys.}\ }\textbf
  {\bibinfo {volume} {116}},\ \bibinfo {pages} {10997--11010} (\bibinfo {year}
  {2002})}\BibitemShut {NoStop}%
\bibitem [{\citenamefont {Mihalcescu}\ \emph {et~al.}(2004)\citenamefont
  {Mihalcescu}, \citenamefont {Hsing},\ and\ \citenamefont
  {Leibler}}]{mihalcescu04}%
  \BibitemOpen
  \bibfield  {author} {\bibinfo {author} {\bibfnamefont {I.}~\bibnamefont
  {Mihalcescu}}, \bibinfo {author} {\bibfnamefont {W.}~\bibnamefont {Hsing}}, \
  and\ \bibinfo {author} {\bibfnamefont {S.}~\bibnamefont {Leibler}},\
  }\bibfield  {title} {\enquote {\bibinfo {title} {Resilient circadian
  oscillator revealed in individual cyanobacteria},}\ }\href@noop {} {\bibfield
   {journal} {\bibinfo  {journal} {Nature}\ }\textbf {\bibinfo {volume}
  {430}},\ \bibinfo {pages} {81--85} (\bibinfo {year} {2004})}\BibitemShut
  {NoStop}%
\bibitem [{\citenamefont {Potoyan}\ and\ \citenamefont
  {Wolynes}(2014)}]{potoyan14}%
  \BibitemOpen
  \bibfield  {author} {\bibinfo {author} {\bibfnamefont {D.~A.}\ \bibnamefont
  {Potoyan}}\ and\ \bibinfo {author} {\bibfnamefont {P.~G.}\ \bibnamefont
  {Wolynes}},\ }\bibfield  {title} {\enquote {\bibinfo {title} {{On the
  dephasing of genetic oscillators}},}\ }\href@noop {} {\bibfield  {journal}
  {\bibinfo  {journal} {Proc. Natl. Acad. Sci. USA}\ }\textbf {\bibinfo
  {volume} {111}},\ \bibinfo {pages} {2391--2396} (\bibinfo {year}
  {2014})}\BibitemShut {NoStop}%
\bibitem [{\citenamefont {Webb}\ \emph {et~al.}(2016)\citenamefont {Webb},
  \citenamefont {Lengyel}, \citenamefont {J{\"o}rg}, \citenamefont {Valentin},
  \citenamefont {J{\"u}licher}, \citenamefont {Morelli},\ and\ \citenamefont
  {Oates}}]{webb16}%
  \BibitemOpen
  \bibfield  {author} {\bibinfo {author} {\bibfnamefont {A.~B.}\ \bibnamefont
  {Webb}}, \bibinfo {author} {\bibfnamefont {I.~M.}\ \bibnamefont {Lengyel}},
  \bibinfo {author} {\bibfnamefont {D.~J.}\ \bibnamefont {J{\"o}rg}}, \bibinfo
  {author} {\bibfnamefont {G.}~\bibnamefont {Valentin}}, \bibinfo {author}
  {\bibfnamefont {F.}~\bibnamefont {J{\"u}licher}}, \bibinfo {author}
  {\bibfnamefont {L.~G.}\ \bibnamefont {Morelli}}, \ and\ \bibinfo {author}
  {\bibfnamefont {A.~C.}\ \bibnamefont {Oates}},\ }\bibfield  {title} {\enquote
  {\bibinfo {title} {{Persistence, period and precision of autonomous cellular
  oscillators from the zebrafish segmentation clock}},}\ }\href@noop {}
  {\bibfield  {journal} {\bibinfo  {journal} {eLife}\ }\textbf {\bibinfo
  {volume} {5}},\ \bibinfo {pages} {e08438} (\bibinfo {year}
  {2016})}\BibitemShut {NoStop}%
\bibitem [{\citenamefont {Lengyel}\ and\ \citenamefont
  {Morelli}(2017)}]{lengyel17}%
  \BibitemOpen
  \bibfield  {author} {\bibinfo {author} {\bibfnamefont {I.~M.}\ \bibnamefont
  {Lengyel}}\ and\ \bibinfo {author} {\bibfnamefont {L.~G.}\ \bibnamefont
  {Morelli}},\ }\bibfield  {title} {\enquote {\bibinfo {title} {{Multiple
  binding sites for transcriptional repressors can produce regular bursting and
  enhance noise suppression}},}\ }\href@noop {} {\bibfield  {journal} {\bibinfo
   {journal} {Phys. Rev. E}\ }\textbf {\bibinfo {volume} {95}},\ \bibinfo
  {pages} {042412} (\bibinfo {year} {2017})}\BibitemShut {NoStop}%
\bibitem [{\citenamefont {Lewis}(2003)}]{lewis03}%
  \BibitemOpen
  \bibfield  {author} {\bibinfo {author} {\bibfnamefont {J.}~\bibnamefont
  {Lewis}},\ }\bibfield  {title} {\enquote {\bibinfo {title} {Autoinhibition
  with transcriptional delay: A simple mechanism for the zebrafish
  somitogenesis oscillator.}}\ }\href@noop {} {\bibfield  {journal} {\bibinfo
  {journal} {Curr. Biol.}\ }\textbf {\bibinfo {volume} {13}},\ \bibinfo {pages}
  {1398--1408} (\bibinfo {year} {2003})}\BibitemShut {NoStop}%
\bibitem [{\citenamefont {Liu}\ \emph {et~al.}(2007)\citenamefont {Liu},
  \citenamefont {Welsh}, \citenamefont {Ko}, \citenamefont {Tran},
  \citenamefont {Zhang}, \citenamefont {Priest}, \citenamefont {Buhr},
  \citenamefont {Singer}, \citenamefont {Meeker}, \citenamefont {Verma},
  \citenamefont {Doyle~III}, \citenamefont {Takahashi},\ and\ \citenamefont
  {Kay}}]{liu07}%
  \BibitemOpen
  \bibfield  {author} {\bibinfo {author} {\bibfnamefont {A.~C.}\ \bibnamefont
  {Liu}}, \bibinfo {author} {\bibfnamefont {D.~K.}\ \bibnamefont {Welsh}},
  \bibinfo {author} {\bibfnamefont {C.~H.}\ \bibnamefont {Ko}}, \bibinfo
  {author} {\bibfnamefont {H.~G.}\ \bibnamefont {Tran}}, \bibinfo {author}
  {\bibfnamefont {E.~E.}\ \bibnamefont {Zhang}}, \bibinfo {author}
  {\bibfnamefont {A.~A.}\ \bibnamefont {Priest}}, \bibinfo {author}
  {\bibfnamefont {E.~D.}\ \bibnamefont {Buhr}}, \bibinfo {author}
  {\bibfnamefont {O.}~\bibnamefont {Singer}}, \bibinfo {author} {\bibfnamefont
  {K.}~\bibnamefont {Meeker}}, \bibinfo {author} {\bibfnamefont {I.~M.}\
  \bibnamefont {Verma}}, \bibinfo {author} {\bibfnamefont {F.~J.}\ \bibnamefont
  {Doyle~III}}, \bibinfo {author} {\bibfnamefont {J.~S.}\ \bibnamefont
  {Takahashi}}, \ and\ \bibinfo {author} {\bibfnamefont {S.~A.}\ \bibnamefont
  {Kay}},\ }\bibfield  {title} {\enquote {\bibinfo {title} {{Intercellular
  Coupling Confers Robustness against Mutations in the SCN Circadian Clock
  Network}},}\ }\href@noop {} {\bibfield  {journal} {\bibinfo  {journal}
  {Cell}\ }\textbf {\bibinfo {volume} {129}},\ \bibinfo {pages} {605--616}
  (\bibinfo {year} {2007})}\BibitemShut {NoStop}%
\bibitem [{\citenamefont {Needleman}\ \emph {et~al.}(2001)\citenamefont
  {Needleman}, \citenamefont {Tiesinga},\ and\ \citenamefont
  {Sejnowski}}]{needleman01}%
  \BibitemOpen
  \bibfield  {author} {\bibinfo {author} {\bibfnamefont {D.~J.}\ \bibnamefont
  {Needleman}}, \bibinfo {author} {\bibfnamefont {P.~H.~E.}\ \bibnamefont
  {Tiesinga}}, \ and\ \bibinfo {author} {\bibfnamefont {T.~J.}\ \bibnamefont
  {Sejnowski}},\ }\bibfield  {title} {\enquote {\bibinfo {title} {{Collective
  enhancement of precision in networks of coupled oscillators}},}\ }\href@noop
  {} {\bibfield  {journal} {\bibinfo  {journal} {Physica D: Nonlinear
  Phenomena}\ }\textbf {\bibinfo {volume} {155}},\ \bibinfo {pages} {324--336}
  (\bibinfo {year} {2001})}\BibitemShut {NoStop}%
\bibitem [{\citenamefont {Morelli}\ \emph {et~al.}(2009)\citenamefont
  {Morelli}, \citenamefont {Ares}, \citenamefont {Herrgen}, \citenamefont
  {Schr{\"{o}}ter}, \citenamefont {J{\"{u}}licher},\ and\ \citenamefont
  {Oates}}]{morelli09}%
  \BibitemOpen
  \bibfield  {author} {\bibinfo {author} {\bibfnamefont {L.~G.}\ \bibnamefont
  {Morelli}}, \bibinfo {author} {\bibfnamefont {S.}~\bibnamefont {Ares}},
  \bibinfo {author} {\bibfnamefont {L.}~\bibnamefont {Herrgen}}, \bibinfo
  {author} {\bibfnamefont {C.}~\bibnamefont {Schr{\"{o}}ter}}, \bibinfo
  {author} {\bibfnamefont {F.}~\bibnamefont {J{\"{u}}licher}}, \ and\ \bibinfo
  {author} {\bibfnamefont {A.~C.}\ \bibnamefont {Oates}},\ }\bibfield  {title}
  {\enquote {\bibinfo {title} {Delayed coupling theory of vertebrate
  segmentation},}\ }\href@noop {} {\bibfield  {journal} {\bibinfo  {journal}
  {HFSP J.}\ }\textbf {\bibinfo {volume} {3}},\ \bibinfo {pages} {55--66}
  (\bibinfo {year} {2009})}\BibitemShut {NoStop}%
\bibitem [{\citenamefont {Herrgen}\ \emph {et~al.}(2010)\citenamefont
  {Herrgen}, \citenamefont {Ares}, \citenamefont {Morelli}, \citenamefont
  {Schr{\"{o}}ter}, \citenamefont {J{\"{u}}licher},\ and\ \citenamefont
  {Oates}}]{herrgen10}%
  \BibitemOpen
  \bibfield  {author} {\bibinfo {author} {\bibfnamefont {L.}~\bibnamefont
  {Herrgen}}, \bibinfo {author} {\bibfnamefont {S.}~\bibnamefont {Ares}},
  \bibinfo {author} {\bibfnamefont {L.~G.}\ \bibnamefont {Morelli}}, \bibinfo
  {author} {\bibfnamefont {C.}~\bibnamefont {Schr{\"{o}}ter}}, \bibinfo
  {author} {\bibfnamefont {F.}~\bibnamefont {J{\"{u}}licher}}, \ and\ \bibinfo
  {author} {\bibfnamefont {A.~C.}\ \bibnamefont {Oates}},\ }\bibfield  {title}
  {\enquote {\bibinfo {title} {Intercellular coupling regulates the period of
  the segmentation clock},}\ }\href@noop {} {\bibfield  {journal} {\bibinfo
  {journal} {Curr. Biol.}\ }\textbf {\bibinfo {volume} {20}},\ \bibinfo {pages}
  {1244--1253} (\bibinfo {year} {2010})}\BibitemShut {NoStop}%
\bibitem [{\citenamefont {Ares}\ \emph {et~al.}(2012)\citenamefont {Ares},
  \citenamefont {Morelli}, \citenamefont {J{\"o}rg}, \citenamefont {Oates},\
  and\ \citenamefont {J{\"u}licher}}]{ares12}%
  \BibitemOpen
  \bibfield  {author} {\bibinfo {author} {\bibfnamefont {S.}~\bibnamefont
  {Ares}}, \bibinfo {author} {\bibfnamefont {L.~G.}\ \bibnamefont {Morelli}},
  \bibinfo {author} {\bibfnamefont {D.~J.}\ \bibnamefont {J{\"o}rg}}, \bibinfo
  {author} {\bibfnamefont {A.~C.}\ \bibnamefont {Oates}}, \ and\ \bibinfo
  {author} {\bibfnamefont {F.}~\bibnamefont {J{\"u}licher}},\ }\bibfield
  {title} {\enquote {\bibinfo {title} {Collective modes of coupled phase
  oscillators with delayed coupling},}\ }\href@noop {} {\bibfield  {journal}
  {\bibinfo  {journal} {Phys. Rev. Lett.}\ }\textbf {\bibinfo {volume} {108}},\
  \bibinfo {pages} {204101} (\bibinfo {year} {2012})}\BibitemShut {NoStop}%
\bibitem [{\citenamefont {Cross}(2012)}]{cross12}%
  \BibitemOpen
  \bibfield  {author} {\bibinfo {author} {\bibfnamefont {M.~C.}\ \bibnamefont
  {Cross}},\ }\bibfield  {title} {\enquote {\bibinfo {title} {{Improving the
  frequency precision of oscillators by synchronization}},}\ }\href@noop {}
  {\bibfield  {journal} {\bibinfo  {journal} {Phys. Rev. E}\ }\textbf {\bibinfo
  {volume} {85}},\ \bibinfo {pages} {046214} (\bibinfo {year}
  {2012})}\BibitemShut {NoStop}%
\bibitem [{\citenamefont {Shimojo}\ \emph {et~al.}(2016)\citenamefont
  {Shimojo}, \citenamefont {Isomura}, \citenamefont {Ohtsuka}, \citenamefont
  {Kori}, \citenamefont {Miyachi},\ and\ \citenamefont {Kageyama}}]{shimojo16}%
  \BibitemOpen
  \bibfield  {author} {\bibinfo {author} {\bibfnamefont {H.}~\bibnamefont
  {Shimojo}}, \bibinfo {author} {\bibfnamefont {A.}~\bibnamefont {Isomura}},
  \bibinfo {author} {\bibfnamefont {T.}~\bibnamefont {Ohtsuka}}, \bibinfo
  {author} {\bibfnamefont {H.}~\bibnamefont {Kori}}, \bibinfo {author}
  {\bibfnamefont {H.}~\bibnamefont {Miyachi}}, \ and\ \bibinfo {author}
  {\bibfnamefont {R.}~\bibnamefont {Kageyama}},\ }\bibfield  {title} {\enquote
  {\bibinfo {title} {{Oscillatory control of Delta-like1 in cell interactions
  regulates dynamic gene expression and tissue morphogenesis}},}\ }\href@noop
  {} {\bibfield  {journal} {\bibinfo  {journal} {Genes Dev.}\ }\textbf
  {\bibinfo {volume} {30}},\ \bibinfo {pages} {102--116} (\bibinfo {year}
  {2016})}\BibitemShut {NoStop}%
\bibitem [{\citenamefont {Liao}\ \emph {et~al.}(2016)\citenamefont {Liao},
  \citenamefont {J{\"o}rg},\ and\ \citenamefont {Oates}}]{liao16}%
  \BibitemOpen
  \bibfield  {author} {\bibinfo {author} {\bibfnamefont {B.-K.}\ \bibnamefont
  {Liao}}, \bibinfo {author} {\bibfnamefont {D.~J.}\ \bibnamefont {J{\"o}rg}},
  \ and\ \bibinfo {author} {\bibfnamefont {A.~C.}\ \bibnamefont {Oates}},\
  }\bibfield  {title} {\enquote {\bibinfo {title} {{Faster embryonic
  segmentation through elevated Delta-Notch signaling}},}\ }\href@noop {}
  {\bibfield  {journal} {\bibinfo  {journal} {Nat. Commun.}\ }\textbf {\bibinfo
  {volume} {7}},\ \bibinfo {pages} {11861} (\bibinfo {year}
  {2016})}\BibitemShut {NoStop}%
\bibitem [{\citenamefont {J\"org}(2017)}]{jorg17}%
  \BibitemOpen
  \bibfield  {author} {\bibinfo {author} {\bibfnamefont {D.~J.}\ \bibnamefont
  {J\"org}},\ }\bibfield  {title} {\enquote {\bibinfo {title} {{Stochastic
  Kuramoto oscillators with discrete phase states}},}\ }\href@noop {}
  {\bibfield  {journal} {\bibinfo  {journal} {Phys. Rev. E}\ }\textbf {\bibinfo
  {volume} {96}},\ \bibinfo {pages} {032201} (\bibinfo {year}
  {2017})}\BibitemShut {NoStop}%
\bibitem [{\citenamefont {Isomura}\ \emph {et~al.}(2017)\citenamefont
  {Isomura}, \citenamefont {Ogushi}, \citenamefont {Kori},\ and\ \citenamefont
  {Kageyama}}]{isomura17}%
  \BibitemOpen
  \bibfield  {author} {\bibinfo {author} {\bibfnamefont {A.}~\bibnamefont
  {Isomura}}, \bibinfo {author} {\bibfnamefont {F.}~\bibnamefont {Ogushi}},
  \bibinfo {author} {\bibfnamefont {H.}~\bibnamefont {Kori}}, \ and\ \bibinfo
  {author} {\bibfnamefont {R.}~\bibnamefont {Kageyama}},\ }\bibfield  {title}
  {\enquote {\bibinfo {title} {{Optogenetic perturbation and bioluminescence
  imaging to analyze cell-to-cell transfer of oscillatory information.}}}\
  }\href@noop {} {\bibfield  {journal} {\bibinfo  {journal} {Genes Dev.}\
  }\textbf {\bibinfo {volume} {31}},\ \bibinfo {pages} {524--535} (\bibinfo
  {year} {2017})}\BibitemShut {NoStop}%
\bibitem [{\citenamefont {Ananthasubramaniam}\ \emph
  {et~al.}(2014)\citenamefont {Ananthasubramaniam}, \citenamefont {Herzog},\
  and\ \citenamefont {Herzel}}]{ananthasubramaniam14}%
  \BibitemOpen
  \bibfield  {author} {\bibinfo {author} {\bibfnamefont {B.}~\bibnamefont
  {Ananthasubramaniam}}, \bibinfo {author} {\bibfnamefont {E.~D.}\ \bibnamefont
  {Herzog}}, \ and\ \bibinfo {author} {\bibfnamefont {H.}~\bibnamefont
  {Herzel}},\ }\bibfield  {title} {\enquote {\bibinfo {title} {{Timing of
  neuropeptide coupling determines synchrony and entrainment in the mammalian
  circadian clock.}}}\ }\href@noop {} {\bibfield  {journal} {\bibinfo
  {journal} {PLoS Comp. Biol.}\ }\textbf {\bibinfo {volume} {10}},\ \bibinfo
  {pages} {e1003565} (\bibinfo {year} {2014})}\BibitemShut {NoStop}%
\bibitem [{\citenamefont {Lewis}\ \emph {et~al.}(2009)\citenamefont {Lewis},
  \citenamefont {Hanisch},\ and\ \citenamefont {Holder}}]{lewis09}%
  \BibitemOpen
  \bibfield  {author} {\bibinfo {author} {\bibfnamefont {J.}~\bibnamefont
  {Lewis}}, \bibinfo {author} {\bibfnamefont {A.}~\bibnamefont {Hanisch}}, \
  and\ \bibinfo {author} {\bibfnamefont {M.}~\bibnamefont {Holder}},\
  }\bibfield  {title} {\enquote {\bibinfo {title} {Notch signaling, the
  segmentation clock, and the patterning of vertebrate somites},}\ }\href@noop
  {} {\bibfield  {journal} {\bibinfo  {journal} {J. Biol.}\ }\textbf {\bibinfo
  {volume} {8}},\ \bibinfo {pages} {44} (\bibinfo {year} {2009})}\BibitemShut
  {NoStop}%
\bibitem [{\citenamefont {Gardiner}(2009)}]{gardiner2009}%
  \BibitemOpen
  \bibfield  {author} {\bibinfo {author} {\bibfnamefont {C.}~\bibnamefont
  {Gardiner}},\ }\href {http://books.google.de/books?id=otg3PQAACAAJ} {\emph
  {\bibinfo {title} {{Stochastic Methods: A Handbook for the Natural and Social
  Sciences}}}},\ Springer Series in Synergetics\ (\bibinfo  {publisher}
  {Springer},\ \bibinfo {year} {2009})\BibitemShut {NoStop}%
\bibitem [{\citenamefont {van Kampen}(2007)}]{vanKampen07}%
  \BibitemOpen
  \bibfield  {author} {\bibinfo {author} {\bibfnamefont {N.~G.}\ \bibnamefont
  {van Kampen}},\ }\href@noop {} {\emph {\bibinfo {title} {{Stochastic
  Processes in Physics and Chemistry}}}},\ \bibinfo {edition} {3rd}\ ed.\
  (\bibinfo  {publisher} {Elsevier},\ \bibinfo {year} {2007})\BibitemShut
  {NoStop}%
\bibitem [{\citenamefont {Stratonovich}(1963)}]{stratonovich63}%
  \BibitemOpen
  \bibfield  {author} {\bibinfo {author} {\bibfnamefont {R.~L.}\ \bibnamefont
  {Stratonovich}},\ }\href {http://books.google.de/books?id=vKkOAAAAQAAJ}
  {\emph {\bibinfo {title} {{Topics in the Theory of Random Noise}}}},\
  \bibinfo {series} {Mathematics and its applications}\ No.\ \bibinfo {number}
  {Bd. 1}\ (\bibinfo  {publisher} {Gordon and Breach},\ \bibinfo {year}
  {1963})\BibitemShut {NoStop}%
\bibitem [{\citenamefont {Schuster}\ and\ \citenamefont
  {Wagner}(1989)}]{schuster89}%
  \BibitemOpen
  \bibfield  {author} {\bibinfo {author} {\bibfnamefont {H.~G.}\ \bibnamefont
  {Schuster}}\ and\ \bibinfo {author} {\bibfnamefont {P.}~\bibnamefont
  {Wagner}},\ }\bibfield  {title} {\enquote {\bibinfo {title} {Mutual
  entrainment of two limit cycle oscillators with time delayed coupling.}}\
  }\href@noop {} {\bibfield  {journal} {\bibinfo  {journal} {Prog. Theor.
  Phys.}\ }\textbf {\bibinfo {volume} {81}},\ \bibinfo {pages} {939--945}
  (\bibinfo {year} {1989})}\BibitemShut {NoStop}%
\bibitem [{\citenamefont {Yeung}\ and\ \citenamefont
  {Strogatz}(1999)}]{yeung99}%
  \BibitemOpen
  \bibfield  {author} {\bibinfo {author} {\bibfnamefont {M.~K.~S.}\
  \bibnamefont {Yeung}}\ and\ \bibinfo {author} {\bibfnamefont {S.~H.}\
  \bibnamefont {Strogatz}},\ }\bibfield  {title} {\enquote {\bibinfo {title}
  {Time delay in the {Kuramoto} model of coupled oscillators.}}\ }\href@noop {}
  {\bibfield  {journal} {\bibinfo  {journal} {Phys. Rev. Lett.}\ }\textbf
  {\bibinfo {volume} {82}},\ \bibinfo {pages} {648--651} (\bibinfo {year}
  {1999})}\BibitemShut {NoStop}%
\bibitem [{\citenamefont {Kuramoto}(1984)}]{kuramoto84}%
  \BibitemOpen
  \bibfield  {author} {\bibinfo {author} {\bibfnamefont {Y.}~\bibnamefont
  {Kuramoto}},\ }\bibfield  {title} {\enquote {\bibinfo {title} {Cooperative
  dynamics of oscillator community: a study based on lattice of rings},}\
  }\href@noop {} {\bibfield  {journal} {\bibinfo  {journal} {Prog. Theor.
  Phys.}\ }\textbf {\bibinfo {volume} {79}},\ \bibinfo {pages} {223--240}
  (\bibinfo {year} {1984})}\BibitemShut {NoStop}%
\bibitem [{\citenamefont {Acebr\'{o}n}\ \emph {et~al.}(2005)\citenamefont
  {Acebr\'{o}n}, \citenamefont {Bonilla},\ and\ \citenamefont
  {Vicente}}]{acebron05}%
  \BibitemOpen
  \bibfield  {author} {\bibinfo {author} {\bibfnamefont {J.~A.}\ \bibnamefont
  {Acebr\'{o}n}}, \bibinfo {author} {\bibfnamefont {L.~L.}\ \bibnamefont
  {Bonilla}}, \ and\ \bibinfo {author} {\bibfnamefont {C.~J.~P.}\ \bibnamefont
  {Vicente}},\ }\bibfield  {title} {\enquote {\bibinfo {title} {{The Kuramoto
  model: A simple paradigm for synchronization phenomena}},}\ }\href@noop {}
  {\bibfield  {journal} {\bibinfo  {journal} {Rev. Mod. Phys.}\ }\textbf
  {\bibinfo {volume} {77}},\ \bibinfo {pages} {137} (\bibinfo {year}
  {2005})}\BibitemShut {NoStop}%
\bibitem [{\citenamefont {Rodrigues}\ \emph {et~al.}(2016)\citenamefont
  {Rodrigues}, \citenamefont {Peron}, \citenamefont {Ji},\ and\ \citenamefont
  {Kurths}}]{rodrigues16}%
  \BibitemOpen
  \bibfield  {author} {\bibinfo {author} {\bibfnamefont {F.~A.}\ \bibnamefont
  {Rodrigues}}, \bibinfo {author} {\bibfnamefont {T.~K.~DM.}\ \bibnamefont
  {Peron}}, \bibinfo {author} {\bibfnamefont {P.}~\bibnamefont {Ji}}, \ and\
  \bibinfo {author} {\bibfnamefont {J.}~\bibnamefont {Kurths}},\ }\bibfield
  {title} {\enquote {\bibinfo {title} {{The Kuramoto model in complex
  networks}},}\ }\href@noop {} {\bibfield  {journal} {\bibinfo  {journal}
  {Phys. Rep.}\ }\textbf {\bibinfo {volume} {610}},\ \bibinfo {pages} {1--98}
  (\bibinfo {year} {2016})}\BibitemShut {NoStop}%
\bibitem [{\citenamefont {Earl}\ and\ \citenamefont {Strogatz}(2003)}]{earl03}%
  \BibitemOpen
  \bibfield  {author} {\bibinfo {author} {\bibfnamefont {M.~G.}\ \bibnamefont
  {Earl}}\ and\ \bibinfo {author} {\bibfnamefont {S.~H.}\ \bibnamefont
  {Strogatz}},\ }\bibfield  {title} {\enquote {\bibinfo {title}
  {Synchronization in oscillator networks with delayed coupling: A stability
  criterion.}}\ }\href@noop {} {\bibfield  {journal} {\bibinfo  {journal}
  {Phys. Rev. E}\ }\textbf {\bibinfo {volume} {67}},\ \bibinfo {pages} {036204}
  (\bibinfo {year} {2003})}\BibitemShut {NoStop}%
\bibitem [{Note1()}]{Note1}%
  \BibitemOpen
  \bibinfo {note} {The logarithmic periodogram $\protect \mathcal {P}$ of a
  time series $(x_1,\mathinner {\ldotp \ldotp \ldotp },x_m)$ is given by
  $\protect \mathcal {P}_\omega =2 \protect \qopname \relax o{ln}|\protect
  \mathaccentV {hat}05Ex_\omega |$, where $\protect \mathaccentV
  {hat}05Ex_\omega $ is the discrete Fourier transform of $x_k$.}\BibitemShut
  {Stop}%
\bibitem [{Note2()}]{Note2}%
  \BibitemOpen
  \bibinfo {note} {Note that the phase model Eq.~({\ref
  {stos.phaseoscillators}}) is an infinite-dimensional system due to the
  presence of delays and therefore requires an entire phase history $\phi
  _i(t)|_{t\leq 0}$ as an initial condition \unhbox \voidb@x \hbox {\cite
  {schuster89}}. Therefore, the final state of the system may depend on the
  entire time dependence of the initial history. For simplicity, we here only
  focus on constant initial conditions.}\BibitemShut {Stop}%
\bibitem [{\citenamefont {Ao}\ \emph {et~al.}(2013)\citenamefont {Ao},
  \citenamefont {H\"{a}nggi},\ and\ \citenamefont {Schmid}}]{ao13}%
  \BibitemOpen
  \bibfield  {author} {\bibinfo {author} {\bibfnamefont {X.}~\bibnamefont
  {Ao}}, \bibinfo {author} {\bibfnamefont {P.}~\bibnamefont {H\"{a}nggi}}, \
  and\ \bibinfo {author} {\bibfnamefont {G.}~\bibnamefont {Schmid}},\
  }\bibfield  {title} {\enquote {\bibinfo {title} {{In-phase and anti-phase
  synchronization in noisy Hodgkin-Huxley neurons}},}\ }\href@noop {}
  {\bibfield  {journal} {\bibinfo  {journal} {Math. Biosci.}\ }\textbf
  {\bibinfo {volume} {245}},\ \bibinfo {pages} {49--55} (\bibinfo {year}
  {2013})}\BibitemShut {NoStop}%
\bibitem [{\citenamefont {D'Huys}\ \emph {et~al.}(2014)\citenamefont {D'Huys},
  \citenamefont {J{\"u}ngling},\ and\ \citenamefont {Kinzel}}]{dhuys14}%
  \BibitemOpen
  \bibfield  {author} {\bibinfo {author} {\bibfnamefont {O.}~\bibnamefont
  {D'Huys}}, \bibinfo {author} {\bibfnamefont {Th.}\ \bibnamefont
  {J{\"u}ngling}}, \ and\ \bibinfo {author} {\bibfnamefont {W.}~\bibnamefont
  {Kinzel}},\ }\bibfield  {title} {\enquote {\bibinfo {title} {{Stochastic
  switching in delay-coupled oscillators}},}\ }\href@noop {} {\bibfield
  {journal} {\bibinfo  {journal} {Phys. Rev. E}\ }\textbf {\bibinfo {volume}
  {90}},\ \bibinfo {pages} {032918--9} (\bibinfo {year} {2014})}\BibitemShut
  {NoStop}%
\bibitem [{\citenamefont {Matsuda}\ \emph {et~al.}(2015)\citenamefont
  {Matsuda}, \citenamefont {Koga}, \citenamefont {Woltjen}, \citenamefont
  {Nishida},\ and\ \citenamefont {Ebisuya}}]{matsuda15}%
  \BibitemOpen
  \bibfield  {author} {\bibinfo {author} {\bibfnamefont {M.}~\bibnamefont
  {Matsuda}}, \bibinfo {author} {\bibfnamefont {M.}~\bibnamefont {Koga}},
  \bibinfo {author} {\bibfnamefont {K.}~\bibnamefont {Woltjen}}, \bibinfo
  {author} {\bibfnamefont {E.}~\bibnamefont {Nishida}}, \ and\ \bibinfo
  {author} {\bibfnamefont {M.}~\bibnamefont {Ebisuya}},\ }\bibfield  {title}
  {\enquote {\bibinfo {title} {Synthetic lateral inhibition governs cell-type
  bifurcation with robust ratios},}\ }\href@noop {} {\bibfield  {journal}
  {\bibinfo  {journal} {Nat. Commun.}\ }\textbf {\bibinfo {volume} {6}},\
  \bibinfo {pages} {6195} (\bibinfo {year} {2015})}\BibitemShut {NoStop}%
\bibitem [{\citenamefont {Schr{\"{o}}ter}\ \emph {et~al.}(2012)\citenamefont
  {Schr{\"{o}}ter}, \citenamefont {Ares}, \citenamefont {Morelli},
  \citenamefont {Isakova}, \citenamefont {Hens}, \citenamefont {Soroldoni},
  \citenamefont {Gajewski}, \citenamefont {J{\"{u}}licher}, \citenamefont
  {Maerkl}, \citenamefont {Deplancke},\ and\ \citenamefont
  {Oates}}]{schroter12}%
  \BibitemOpen
  \bibfield  {author} {\bibinfo {author} {\bibfnamefont {C.}~\bibnamefont
  {Schr{\"{o}}ter}}, \bibinfo {author} {\bibfnamefont {S.}~\bibnamefont
  {Ares}}, \bibinfo {author} {\bibfnamefont {L.~G.}\ \bibnamefont {Morelli}},
  \bibinfo {author} {\bibfnamefont {A.}~\bibnamefont {Isakova}}, \bibinfo
  {author} {\bibfnamefont {K.}~\bibnamefont {Hens}}, \bibinfo {author}
  {\bibfnamefont {D.}~\bibnamefont {Soroldoni}}, \bibinfo {author}
  {\bibfnamefont {M.}~\bibnamefont {Gajewski}}, \bibinfo {author}
  {\bibfnamefont {F.}~\bibnamefont {J{\"{u}}licher}}, \bibinfo {author}
  {\bibfnamefont {S.~J.}\ \bibnamefont {Maerkl}}, \bibinfo {author}
  {\bibfnamefont {B.}~\bibnamefont {Deplancke}}, \ and\ \bibinfo {author}
  {\bibfnamefont {A.~C.}\ \bibnamefont {Oates}},\ }\bibfield  {title} {\enquote
  {\bibinfo {title} {Topology and dynamics of the zebrafish segmentation clock
  core circuit},}\ }\href@noop {} {\bibfield  {journal} {\bibinfo  {journal}
  {PLoS Biology}\ }\textbf {\bibinfo {volume} {10}},\ \bibinfo {pages}
  {e1001364} (\bibinfo {year} {2012})}\BibitemShut {NoStop}%
\bibitem [{\citenamefont {Bernard}\ \emph {et~al.}(2007)\citenamefont
  {Bernard}, \citenamefont {Gonze}, \citenamefont {\v{C}ajavec}, \citenamefont
  {Herzel},\ and\ \citenamefont {Kramer}}]{bernard07}%
  \BibitemOpen
  \bibfield  {author} {\bibinfo {author} {\bibfnamefont {S.}~\bibnamefont
  {Bernard}}, \bibinfo {author} {\bibfnamefont {D.}~\bibnamefont {Gonze}},
  \bibinfo {author} {\bibfnamefont {B.}~\bibnamefont {\v{C}ajavec}}, \bibinfo
  {author} {\bibfnamefont {H.}~\bibnamefont {Herzel}}, \ and\ \bibinfo {author}
  {\bibfnamefont {A.}~\bibnamefont {Kramer}},\ }\bibfield  {title} {\enquote
  {\bibinfo {title} {{Synchronization-Induced Rhythmicity of Circadian
  Oscillators in the Suprachiasmatic Nucleus}},}\ }\href@noop {} {\bibfield
  {journal} {\bibinfo  {journal} {PLoS Comp. Biol.}\ }\textbf {\bibinfo
  {volume} {3}},\ \bibinfo {pages} {e68} (\bibinfo {year} {2007})}\BibitemShut
  {NoStop}%
\bibitem [{\citenamefont {Azzi}\ \emph {et~al.}(2017)\citenamefont {Azzi},
  \citenamefont {Evans}, \citenamefont {Leise}, \citenamefont {Myung},
  \citenamefont {Takumi}, \citenamefont {Davidson},\ and\ \citenamefont
  {Brown}}]{azzi17}%
  \BibitemOpen
  \bibfield  {author} {\bibinfo {author} {\bibfnamefont {A.}~\bibnamefont
  {Azzi}}, \bibinfo {author} {\bibfnamefont {J.~A.}\ \bibnamefont {Evans}},
  \bibinfo {author} {\bibfnamefont {T.}~\bibnamefont {Leise}}, \bibinfo
  {author} {\bibfnamefont {J.}~\bibnamefont {Myung}}, \bibinfo {author}
  {\bibfnamefont {T.}~\bibnamefont {Takumi}}, \bibinfo {author} {\bibfnamefont
  {A.~J.}\ \bibnamefont {Davidson}}, \ and\ \bibinfo {author} {\bibfnamefont
  {S.~A.}\ \bibnamefont {Brown}},\ }\bibfield  {title} {\enquote {\bibinfo
  {title} {{Network Dynamics Mediate Circadian Clock Plasticity}},}\
  }\href@noop {} {\bibfield  {journal} {\bibinfo  {journal} {Neuron}\ }\textbf
  {\bibinfo {volume} {93}},\ \bibinfo {pages} {441} (\bibinfo {year}
  {2017})}\BibitemShut {NoStop}%
\bibitem [{\citenamefont {Erzberger}\ \emph {et~al.}(2013)\citenamefont
  {Erzberger}, \citenamefont {Hampp}, \citenamefont {Granada}, \citenamefont
  {Albrecht},\ and\ \citenamefont {Herzel}}]{erzberger13}%
  \BibitemOpen
  \bibfield  {author} {\bibinfo {author} {\bibfnamefont {A.}~\bibnamefont
  {Erzberger}}, \bibinfo {author} {\bibfnamefont {G.}~\bibnamefont {Hampp}},
  \bibinfo {author} {\bibfnamefont {A.~E.}\ \bibnamefont {Granada}}, \bibinfo
  {author} {\bibfnamefont {U.}~\bibnamefont {Albrecht}}, \ and\ \bibinfo
  {author} {\bibfnamefont {H.}~\bibnamefont {Herzel}},\ }\bibfield  {title}
  {\enquote {\bibinfo {title} {{Genetic redundancy strengthens the circadian
  clock leading to a narrow entrainment range}},}\ }\href@noop {} {\bibfield
  {journal} {\bibinfo  {journal} {J. Royal Soc. Interface}\ }\textbf {\bibinfo
  {volume} {10}},\ \bibinfo {pages} {20130221} (\bibinfo {year}
  {2013})}\BibitemShut {NoStop}%
\bibitem [{\citenamefont {Gillespie}(1977)}]{gillespie77}%
  \BibitemOpen
  \bibfield  {author} {\bibinfo {author} {\bibfnamefont {D.}~\bibnamefont
  {Gillespie}},\ }\bibfield  {title} {\enquote {\bibinfo {title} {Exact
  stochastic simulation of coupled chemical-reactions},}\ }\href@noop {}
  {\bibfield  {journal} {\bibinfo  {journal} {J. Phys. Chem.}\ }\textbf
  {\bibinfo {volume} {81}},\ \bibinfo {pages} {2340--2361} (\bibinfo {year}
  {1977})}\BibitemShut {NoStop}%
\bibitem [{\citenamefont {Torrence}\ and\ \citenamefont
  {Compo}(1998)}]{torrence98}%
  \BibitemOpen
  \bibfield  {author} {\bibinfo {author} {\bibfnamefont {C.}~\bibnamefont
  {Torrence}}\ and\ \bibinfo {author} {\bibfnamefont {G.~P.}\ \bibnamefont
  {Compo}},\ }\bibfield  {title} {\enquote {\bibinfo {title} {A practical guide
  to wavelet analysis},}\ }\href@noop {} {\bibfield  {journal} {\bibinfo
  {journal} {B. Am. Meteorol. Soc.}\ }\textbf {\bibinfo {volume} {79}},\
  \bibinfo {pages} {61--78} (\bibinfo {year} {1998})}\BibitemShut {NoStop}%
\bibitem [{\citenamefont {Strogatz}(1994)}]{strogatz}%
  \BibitemOpen
  \bibfield  {author} {\bibinfo {author} {\bibfnamefont {S.~H.}\ \bibnamefont
  {Strogatz}},\ }\href@noop {} {\emph {\bibinfo {title} {Nonlinear Dynamics and
  Chaos}}}\ (\bibinfo  {publisher} {Addison-Wesley},\ \bibinfo {address}
  {Reading, MA},\ \bibinfo {year} {1994})\BibitemShut {NoStop}%
\bibitem [{\citenamefont {Amann}\ \emph {et~al.}(2007)\citenamefont {Amann},
  \citenamefont {Sch\"{o}ll},\ and\ \citenamefont {Just}}]{amann07}%
  \BibitemOpen
  \bibfield  {author} {\bibinfo {author} {\bibfnamefont {A.}~\bibnamefont
  {Amann}}, \bibinfo {author} {\bibfnamefont {E.}~\bibnamefont {Sch\"{o}ll}}, \
  and\ \bibinfo {author} {\bibfnamefont {W.}~\bibnamefont {Just}},\ }\bibfield
  {title} {\enquote {\bibinfo {title} {{Some basic remarks on eigenmode
  expansions of time-delay dynamics}},}\ }\href@noop {} {\bibfield  {journal}
  {\bibinfo  {journal} {Physica A}\ }\textbf {\bibinfo {volume} {373}},\
  \bibinfo {pages} {191--202} (\bibinfo {year} {2007})}\BibitemShut {NoStop}%
\end{thebibliography}
\end{document}